\pdfoutput=1
\documentclass[aps,prc,reprint,superscriptaddress,nofootinbib]{revtex4-1}

\usepackage{amsmath,amssymb}    
\usepackage{graphicx}   
\usepackage[dvipsnames]{xcolor} 
\usepackage{subcaption}
\DeclareCaptionJustification{fulljustified}{%
  \leftskip0pt\rightskip0pt\parfillskip0pt plus 1fil\relax}
\usepackage{color}
\usepackage{siunitx}  
\DeclareSIUnit{\MeV}{MeV}
\DeclareSIUnit{\GeV}{GeV}
\DeclareSIUnit{\TeV}{TeV}
\DeclareSIUnit{\fm}{fm}
\usepackage[colorlinks,linkcolor=blue,citecolor=blue]{hyperref}   
\usepackage{bm} 
\usepackage[capitalise]{cleveref}

\def\p{{\bf p}}

\def\f0{f^{(0)}}

\newcommand\nda{\end{align}}
\newcommand{\FastReso}{\textsc{FastReso}}

\def\st{\begin{equation}}
\def\stp{\end{equation}}
\def\bg{\begin{eqnarray}}
\def\nd{\end{eqnarray}}

\advance\parskip 1.9pt
\advance\voffset -0.2in

\newcommand{\rmd}{\mathrm{d}}

\definecolor{oscarC}{RGB}{22, 156, 172}

\begin{document}
\title{Effect of thermal broadening on light hadron production in heavy-ion collisions}
\author{Q'inich Coc}%
    \email{figueroa@thphys.uni-heidelberg.de}
\affiliation{Institut für Theoretische Physik, Universität Heidelberg, 69120 Heidelberg, Germany}
\author{Oscar Garcia-Montero}%
    \email{oscarjesus.garcia@usc.es}
\affiliation{Instituto Galego de F\'isica de Altas Enerx\'ias IGFAE, Universidade de Santiago de Compostela, E-15782 Galicia, Spain}
\author{Aleksas Mazeliauskas}%
    \email{a.mazeliauskas@thphys.uni-heidelberg.de}
\affiliation{Institut für Theoretische Physik, Universität Heidelberg, 69120 Heidelberg, Germany}
\begin{abstract}

The measured hadron yields in heavy-ion collisions are well described by thermal particle production at temperatures close to the QCD pseudo-critical transition.
However, if particles are produced in thermal equilibrium, then the in-medium effects, e.g., broadening of their spectral functions, must be taken into account. In this work, we study for the first time the effect of the thermal broadening of the light vector-meson spectral functions on light hadron production in heavy-ion collisions. 
We generalize the efficient resonance decay framework {\FastReso} to resonances with finite-width spectral functions. We then compute the pion, kaon and proton spectra from the decays of hadron resonances for in-medium and vacuum spectral functions.
We perform a simultaneous blast-wave fit with decay feed-down to measured particle spectra in central Pb-Pb and Xe-Xe collisions at the LHC.
We find an increase in pion spectra at low momenta from including thermally broadened $\rho$ and $\omega$ mesons. Additional pion yield from finite resonance widths reduces the commonly observed discrepancy between models and data, but does not completely resolve the soft pion puzzle in heavy-ion collisions.

\end{abstract}
\maketitle

\section{Introduction}

At sufficiently high temperatures achieved in relativistic heavy-ion collisions, nuclear matter undergoes a transition to a deconfined phase in which quarks and gluons become the relevant dynamical degrees of freedom~\cite{Harris:2023tti}. In this high-temperature phase of QCD, known as the quark–gluon plasma (QGP), chiral symmetry is (approximately) restored~\cite{Rapp:1999ej}.
A comprehensive experimental program has established a remarkably successful description of hot QCD matter as a nearly perfect fluid with quantitative predictions for a wide range of soft hadronic observables~\cite{Heinz:2013th}.  

Despite this success, state-of-the-art viscous-hydrodynamic simulations systematically under-predict the yield of low transverse-momentum pions measured at LHC energies, a long-standing discrepancy commonly referred to as the soft-pion puzzle~\cite{Lu:2024shm}. A number of explanations have been proposed to account for this low-momentum excess, including modifications of the vacuum $\rho$-meson spectral width~\cite{Huovinen:2016xxq}, pion condensation~\cite{Begun:2015ifa,Begun:2016cva}, a perturbative corona contribution~\cite{Kanakubo:2019ogh}, and
changes to the pion dispersion relation~\cite{Fuchs:1996pa}, yet none of these has so far fully resolved the discrepancy. 
It is worth noting that chiral symmetry restoration plays virtually no role in any of these explanations, nor in current hydrodynamic models of QGP expansion more broadly~\cite{Elfner:2022iae}.
Recently, the role of chiral symmetry restoration in direct soft-pion production has been studied in both equilibrium~\cite{Grossi:2021gqi} and out-of-equilibrium~\cite{Florio:2025lvu,Florio:2025zqv,Bruschke:2025wny} settings, and it may offer a resolution to this puzzle. In this work, we investigate how chiral symmetry restoration affects the resonance decay feed-down to soft pions in thermal equilibrium.

One of the prime candidates for a direct experimental signature of chiral symmetry restoration is the in-medium broadening of light vector-meson spectral functions. 
Since light vector mesons also constitute the dominant source of decay pions, with around one third of all decay pions coming from $\rho$ and $\omega$ decays~\cite{Qiu:2013wca}, their thermal broadening offers a natural, and largely unexplored, contribution to the soft-pion puzzle. Notably, $\rho$ meson broadening, consistent with chiral symmetry restoration~\cite{Pisarski:1995xu}, has been observed in measurements of dilepton spectra at RHIC~\cite{PHENIX:2015vek,STAR:2013pwb}.

In this work we investigate how the thermal broadening of the $\rho$, $\omega$ and $\phi$ spectral functions affects light hadron production in nuclear collisions. Using a blast-wave freeze-out surface with thermal hadron distributions at constant freeze-out temperature $T_\text{fo}$ and vanishing chemical potential $\mu_B=0$, we compute pion, kaon and proton spectra, focusing in particular on the low-momentum pion regime. To compute the feed-down from resonance decays with thermal spectral functions, we extend the efficient resonance decay code {\FastReso}~\cite{Mazeliauskas:2018irt} to incorporate finite-width spectral functions. The {\FastReso} framework allows the tabulation of irreducible particle spectra components in the fluid rest frame after the resonance decays. Then the observed particle spectra can be computed by integrating over the freeze-out surface with a Lorentz transformation to the lab frame. {\FastReso} has been used in the prediction of light~\cite{Devetak:2019lsk,Mazeliauskas:2019ifr,Kirchner:2023fsj,Lu:2024shm} and heavy flavor~\cite{Andronic:2021erx,Capellino:2023cxe} hadron spectra at the LHC.
We take the resonance masses, branching ratios and resonance widths from the PDG2016 particle list~\cite{ParticleDataGroup:2016lqr,Alba:2017hhe,Alba:2017mqu}. We perform the fit of blast-wave parameters to the experimental measurements from lead-lead (Pb-Pb) and xenon-xenon (Xe-Xe) collisions in the $0$--$5\%$ centrality class and quantify the residual discrepancy with different assumptions on the resonance spectral functions that include vacuum Breit-Wigner and S-matrix spectral functions~\cite{Dashen:1969ep,Huovinen:2016xxq,Lo:2017ldt, Lo:2017sux}, as well as thermally broadened $\rho$, $\omega$ and $\phi$ spectral functions~\cite{Rapp:1997fs,Rapp:1999us,Rapp:1999qu,Rapp:1997ei,Rapp:2000pe,Haglin:1994xu}.

We note that the effect of the $\rho$-meson spectral function in the S-matrix formalism on the pion spectra has already been studied in Ref.~\cite{Huovinen:2016xxq}, where it was found that the finite width of $\rho$ enhances the pion yield at low transverse momentum relative to the zero-width case (see \cref{sec:smatrix}), providing early evidence that resonance widths may be relevant to the soft-pion puzzle. Here we build on this observation by generalizing the {\FastReso} decay framework to arbitrary finite-width spectral functions, and by simultaneously fitting pion, kaon, and proton spectra to quantify this effect together with the in-medium broadening of light vector mesons.
Furthermore, this investigation, performed in local thermal equilibrium, complements work in Refs.~\cite{Florio:2025lvu,Florio:2025zqv,Bruschke:2025wny} where non-equilibrium signals of transition through the chiral crossover are studied.

The paper is organized as follows. \Cref{sec:Spectral_functions} reviews different approaches to spectral functions. \Cref{sec:FastResoFrameWork} introduces the {\FastReso} framework and its extension to incorporate resonance widths, alongside general formulas for these modifications. We also discuss the simultaneous pion, kaon and proton blast-wave fit with resonance feed-down and spectral widths included, \cref{sec:BlastWaveFits}. \Cref{sec:Results} discusses the fits to LHC data. In \cref{sec:conclusions} we present our conclusions and future outlook. Finally, in \cref{sec:numerical_details} we present technical details of finite-width implementation in {\FastReso} code.

While we use the mostly-minus convention for the metric in this paper, in \cref{sec:FastResoFrameWork} we have given convention-agnostic definitions in order not to clash with notation in the original {\FastReso} paper~\cite{Mazeliauskas:2018irt}.

\section{Spectral functions}
\label{sec:Spectral_functions}
In general, the spectral function $\rho_a$ of a particle $a$ is defined as the imaginary part of the propagator of the underlying degrees of freedom~\cite{Kapusta_Gale_2006},
\begin{equation}
\rho_a(m) = -\frac{2m}{\pi}\,\mathrm{Im}\, D_a(m, \vec p=0)\,.
\label{eq:rho_D}
\end{equation}

In an interacting theory, the propagator is determined exactly by the self-energy $\Sigma(p)$ via $D(p)=\big(p^2 - M_a^2 -\Sigma(p)\big)^{-1}$, yielding a unique spectral function. In practice, however, one typically resorts to simplifications---such as effective models, truncations of the self-energy, or assumed functional forms---which lead to analytical or numerical expressions for $\rho_a(m)$ that differ in the specific details of how the resonance is described.

In the following subsections we present the spectral functions used in the computation of particle spectra.
In all cases, we normalize the spectral functions to unity
\begin{equation}
    \int_{t_a}^\infty\!\!  dm\, \rho_a(m)=1\,,
\end{equation}
where $t_a = M_1+ M_2 + \dots$ is the threshold mass, i.e., the lowest mass of the resonance $a$ allowed to decay by energy conservation. For a decay into particles with vanishing spectral widths, this is the sum of their pole masses. For example, the $\Delta^+$ resonance decays through the channels $\Delta^+\rightarrow p + \pi^0$ and $\Delta^+\rightarrow n + \pi^+$, so that $t_a =M_p+M_{\pi^0}$. For decays into resonances with non-trivial spectral functions, the threshold is computed recursively by summing the mass thresholds of the decay products. The precise recursive definition, including the treatment of narrow resonances, is given in \cref{sec:app_thresholds}. When several decay channels are present, we take the lowest possible threshold for the normalization. 

\begin{figure}
\centering\includegraphics[width=1\linewidth]{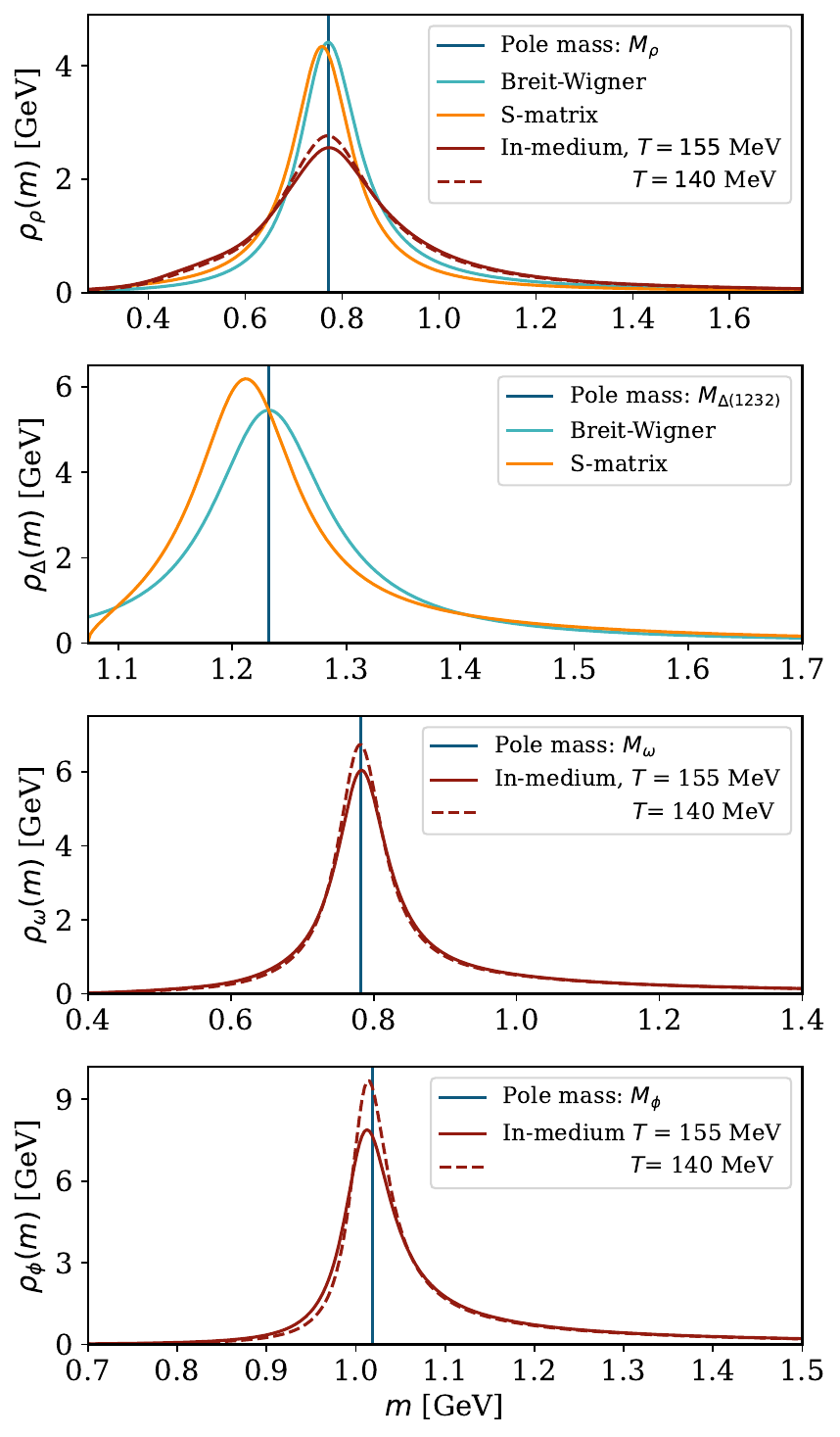}
    \caption{Vacuum and in-medium spectral functions used in this work, shown as a function of mass $m$ for (top to bottom) the $\rho$, $\Delta(1232)$, $\omega$, and $\phi$ resonances. Vertical lines mark the pole masses. For $\rho$ and $\Delta$ we show the vacuum Breit-Wigner (cyan) and S-matrix (orange) parametrizations; for $\rho$, $\omega$, and $\phi$ we additionally show the in-medium spectral functions at $T=\SI{155}{\MeV}$ (solid) and $T=\SI{140}{\MeV}$ (dashed).}
    \label{fig:implemented_spec_funcs}
\end{figure}
\subsection{Vacuum spectral functions}

Even in vacuum, the spectral function encodes how resonances get broadened due to interactions, i.e., instead of a single sharp pole, the state acquires a finite width, and the spectral function gives the probability for the resonance to appear with a different mass as compared to the measured pole mass, $M_a$.

\paragraph{Breit-Wigner parametrization}

The Breit-Wigner form is a parametrization of vacuum spectral functions, where one uses the propagator  
$D(p)=(p^2 - M_a^2 + i\, \sqrt{p^2}\, \Gamma)^{-1}$. This parametrization follows from approximating the self-energy near the pole mass as a constant, $\Sigma(m^2) \simeq - i\, m\, \Gamma$. By introducing this propagator, in the rest frame, into \cref{eq:rho_D}, one gets
\begin{equation}
\label{eq:breitwigner}
\rho_a(m)=\frac{2 \mathcal N_a}{\pi}\frac{m^2\Gamma_a(m)}{(m^2-M_a^2)^2+m^2 \Gamma_a^2(m)}\,, 
\end{equation}
where $\mathcal N_a$ is a normalization factor. The simplest approximation one can take is to assume a constant width $\Gamma_a(m)=\Gamma_a$. In such a case, the width can be taken from the measured value in the particle list used, here PDG2016~\cite{ParticleDataGroup:2016lqr,Alba:2017hhe,Alba:2017mqu}. Additionally, while $\mathcal N_a$ depends on the threshold of the lowest decay channel, the dependence on the cutoff tends to be small and hence normally $\mathcal N_a\approx 1$. In \cref{fig:implemented_spec_funcs}, the reader can find the vacuum (BW) spectral functions for the $\rho$ meson as well as for the $\Delta(1232)$ baryon. Since the vacuum (BW) spectral functions for $\phi$ and $\omega$ are too narrow, i.e., $\Gamma_i/M_i < 0.05$, they are omitted in \cref{fig:implemented_spec_funcs} and are treated as zero-width in the Breit-Wigner and S-matrix scenarios, an approximation which was explicitly found to be valid for the $\omega$-meson in Ref.~\cite{Lo:2017sux}. Their in-medium spectral functions are, however, included in the In-medium scenario, see \cref{sec:inmedium}.

In other frameworks such as the hadronic afterburner SMASH~\cite{Elfner:2025ojd,Hirayama:2022rur,SMASH:2016zqf}, a mass-dependent Breit-Wigner spectral function has been used, where the mass dependence of the total width is obtained by summing the partial widths of all open decay channels, following the Manley-Saleski multichannel parametrization~\cite{Manley:1992yb}. In this method, each partial width is rescaled from its value at the pole mass by the ratio of two-body phase-space factors, expressed in terms of Blatt-Weisskopf functions (which depend on the orbital angular momentum of the decay). 
A more sophisticated case additionally takes into account the finite width of unstable daughter particles, by convolving these phase-space factors with the daughters' own spectral functions, as is done, e.g., for decay chains such as $N^\ast\to\Delta\pi$ in SMASH.

\subsection{S-matrix formulation of the spectral function}\label{sec:smatrix}
In the S-matrix approach \cite{Dashen:1969ep}, the effective spectral function of a resonance is derived from the scattering phase shift $\delta_{IJ}$, where $IJ$ denotes the isospin and spin of the relevant two-body scattering channel. The spectral function is then given by the mass derivative
\begin{equation}
	\rho_a(m)=\frac{1}{\pi} \frac{\rmd \delta_{IJ}}{\rmd m }\,,
\end{equation}
which is automatically normalized to unity when the phase shift rises by $\pi$ across an isolated resonance.

In {\FastReso} we have implemented the S-matrix spectral functions for the $\rho$ meson~\cite{Huovinen:2016xxq} and for the $\Delta(1232)$ baryon~\cite{Lo:2017ldt}. Both are described by the same phenomenological functional form, inspired by a one-loop calculation of the resonance self-energy,
\begin{subequations}
\begin{align}
\delta_{IJ}(m) &= \tan^{-1}\left[-\frac{2}{3m}\frac{\alpha_0}{1+\Pi(q(m))}\frac{q^3(m)}{m^2-M_0^2}\right]\,,\\
\Pi(q) &= c_1 q^2 + c_2 q^4,
\label{eq:smatrix_delta}
\end{align}
\end{subequations}
where $q$ is the center-of-mass momentum of the two decay products $b$ and $c$ in the $a\to b + c$ decay channel
\begin{equation}
q(m)=\frac{1}{2m}\sqrt{\big((m+m_b)^2-m_c^2\big)\big((m-m_b)^2-m_c^2\big)}.
\label{eq:qbc}
\end{equation}
For the $\rho$ meson ($\rho \to \pi + \pi$), we follow Ref.~\cite{Huovinen:2016xxq} and use
$\alpha_0=3.08$, $M_0=\SI{0.77}{\GeV}$, $c_1=\SI{0.59}{\GeV^{-2}}$ and $c_2=0$.
For the $\Delta(1232)$ baryon ($\Delta\to \pi + N$, where $N$ is a proton or neutron), we follow Ref.~\cite{Lo:2017ldt} and use $\alpha_0=45.37$, $M_0=\SI{1.2325}{\GeV}$, $c_1=\SI{16.7}{\GeV^{-2}}$ and $c_2=\SI{65.6}{\GeV^{-4}}$. In computing numerically the spectral function, see \cref{fig:implemented_spec_funcs} for $\rho$ and $\Delta$ we use the pion and nucleon pole masses.

\subsection{In-medium spectral functions}
\label{sec:inmedium}
Extensive lattice QCD calculations have demonstrated that at zero baryon chemical potential QCD matter undergoes a smooth crossover transition from hadronic to partonic degrees of freedom at the pseudo-critical transition temperature $T_\text{pc}\approx \SI{158}{\MeV}$~\cite{Borsanyi:2020fev}. One of the expected manifestations of chiral symmetry restoration is the modification of vector meson spectral functions, where, e.g., light vector mesons $V=\rho,\omega,\phi$ acquire
medium-dependent properties through their interactions with the surrounding thermal medium. As in the vacuum case, these modifications are encoded in the spectral function, \cref{eq:rho_D}. In the case of the medium, we can write the vector meson propagator at rest as
\begin{equation}
    D_V(m,\vec{p}=0;T)
    = (m^2 - m_{V,\mathrm{bare}}^2 - \Sigma_V(m,\vec{p}=0;T))^{-1}.
\end{equation}
Here $\Sigma_V$, the self-energy, receives contributions from hadronic scattering and
from dressing of the meson's dominant decay channels. For the $\rho$ meson,
the coupling to
thermal pions and baryons ($N$, $\Delta$) induces significant
broadening through processes such as $\pi\pi\to\rho$, $\rho N \to B^\ast$~\cite{Rapp:2006xzj}. Additionally, 
baryonic loops ($N\Delta$) soften the pion dispersion relation and
redistribute the spectrum toward lower invariant masses. The narrower
$\omega$ and $\phi$ mesons are modified primarily by collision broadening from
scattering with thermal pions, kaons, and nucleons, as well as by medium
dependence of their dominant decay channels ($\omega\to3\pi$,
$\phi\to K\bar{K}$).

For this phenomenological study we will use thermal spectral functions obtained within hadronic many-body theory, constrained by vacuum properties and the symmetries of the underlying theory, QCD~\cite{Rapp:1997fs}.  Due to the impact of vector mesons on low momentum pions, we will focus on thermal effects within spectral functions of the lightest vector mesons, $\rho$, $\omega$ and $\phi$, computed in Refs.~\cite{Rapp:1999us,Rapp:1999qu,Rapp:1997ei,Rapp:2000pe,Haglin:1994xu} 
for different temperatures. 
While we are aware of studies of  thermal melting of other resonances, such as the $\Delta(1232)$~\cite{vanHees:2004vt,vanHees:2004sv}, $f_2(1270)$, and $a_1(1260)$~\cite{Rapp:2003ar,Hohler:2013eba}, we will leave the extension to higher-mass resonances for later studies.

\section{Particle feed-down with \texorpdfstring{\FastReso}{FastReso}}
\label{sec:FastResoFrameWork}
In this section we briefly introduce the {\FastReso} framework for resonance decays of particles with spectral functions. The details for the case of on-shell resonances can be found in the original paper~\cite{Mazeliauskas:2018irt}. 

\subsection{Decay map $a\to b + X$}

In hydrodynamic models of heavy-ion collisions, the evolution of the dense QCD medium is characterized by fluid fields---such as energy density and the four-velocity $u^\mu$---until microscopic interactions are no longer frequent enough to maintain local thermal equilibrium. At this stage, typically defined by a freeze-out hypersurface $\mathcal{S}$, the fluid fields are mapped onto discrete particles via a particlization prescription that ensures the conservation of energy and momentum flux~\cite{Cooper:1974mv}. 

The subsequent evolution of these particles is governed by the kinetic theory of an interacting hadronic gas~\cite{Elfner:2022iae}. While particle rescatterings continue to modify distributions~\cite{Bass:2000ib,Song:2011qa,Garcia-Montero:2021haa}, the most significant influence on pion, kaon, and proton spectra during the hadronic phase is the decay feed-down from higher-mass resonance states. If one neglects rescatterings, the momentum distribution of a final-state particle $b$ due to the decay feed-down from a parent resonance $a$ is given by the modified Cooper-Frye formula:
\begin{align}
\label{eq:CooperFrye_decaymap}
E_{\mathbf{p}} \frac{dN_b}{d^3\mathbf{p}} = \int dm_a \rho_a(m_a) &\int \frac{d^3\mathbf{q}}{(2\pi)^3 2E_{\mathbf{q}}} D^a_b (p, q)\nonumber\\
\times &\frac{\nu_a}{(2\pi)^3}\int_{\mathcal S}  d\sigma_\mu q^\mu f_a(x,q).
\end{align}
Here, $\rho_a(m_a)$ denotes the spectral function of the parent particle $a$, and $E_{\mathbf{q}} = \sqrt{m_a^2 + \mathbf{q}^2}$ is the on-shell energy. The term $D^a_b(p, q)$ represents the Lorentz-invariant decay map, which encodes the kinematics of the $a \to b + X$ transition. The integration over the freeze-out hypersurface $\mathcal S$ involves the surface element $d\sigma_\mu$ and the local distribution function $f_a(x,q)$, which is determined by the hydrodynamic fluid fields at the particlization point $x^\mu$. Finally, $\nu_a$ accounts for the spin-isospin degeneracy of the parent species.

For an isotropic 2-body decay $a\to b + c$, the decay can be written as~\cite{Sollfrank:1990qz,Byckling:1971vca,ParticleDataGroup:2024cfk}
\begin{equation}
D^a_{b|c}(p^\mu q_\mu) = \int dm_c\rho_c (m_c) B \frac{4\pi^2 m_a}{p^a_{b|c}} \delta(|q^\mu p_\mu| - m_a E^a_{b|c}),
\label{eq:decay_map_dirac}
\end{equation}
where $B$ is the total branching ratio for this process, which is measured at the pole of the resonant peak~\cite{ParticleDataGroup:2016lqr}.

In the rest frame of the decaying particle $a$, the outgoing particle's momentum is fixed by energy-momentum conservation, $|\mathbf{p}| = p^a_{b|c}$, where $p^a_{b|c}=q(m_a)$ (see \cref{eq:qbc}) is the center-of-mass momentum of the decay products evaluated at the parent mass $m_a$.

The particle's energy in this frame is given by
\begin{equation}
E^a_{b|c} \equiv \sqrt{m_b^2 + (p^a_{b|c})^2}.
\end{equation}

The isotropic 3-body decay $a\to b+c+d$ can be treated as a combination of two 2-body decays $a\to b + \tilde c$ and $\tilde c \to c +d$, where the fictitious particle $\tilde c$ has mass $m_{\tilde{c}}^2 = |(p_c + p_d)^2|$~\cite{Sollfrank:1990qz,Byckling:1971vca,ParticleDataGroup:2024cfk}. Then 
\begin{align}
D^a_{b|cd}(p^\mu q_\mu) &=\int dm_c\rho_c (m_c)\int dm_d\rho_d (m_d)\nonumber\\
&\times \frac{\int_{m_c + m_d}^{m_a - m_b} dm_{\tilde{c}} \, p^a_{b|\tilde{c}} p^{\tilde{c}}_{c|d} D^a_{b|\tilde{c}}(p^\mu q_\mu)}
{\int_{m_c + m_d}^{m_a - m_b} dm_{\tilde{c}} \, p^a_{b|\tilde{c}} p^{\tilde{c}}_{c|d}}.\label{eq:3body}
\end{align}
Here $p^{a}_{b|c}$ represents the momentum of particle $b$ in the rest frame of particle $a$, and $p^{\tilde{c}}_{c|d}$ the momentum of particle $c$ or $d$ in their common rest frame~\cite{ParticleDataGroup:2024cfk}. 

More complicated decay chains can be constructed by repeated convolution of 2-body and 3-body decay maps.

\subsection{Irreducible decay functions}

The key idea of the {\FastReso} framework is to reverse the order of the surface integral and the linear map in \cref{eq:CooperFrye_decaymap}. That is, one computes the resonance feed-down from a given surface element before performing the surface integral to get the final particle spectra.
The decays modify the initial $p^\mu f_a$ distribution to a vector distribution function $g^\mu_b$. 
\begin{align}
E_{\mathbf{p}} \frac{dN_b}{d^3\mathbf{p}} =\frac{\nu_b}{(2\pi)^3}\int_{\mathcal S}  d\sigma_\mu g^\mu_b(x,p).\label{eq:freezeout_w_decays}
\end{align}
In the context of viscous hydrodynamics, the particle distribution function on the freeze-out surface depends on the fluid temperature $T$, fluid velocity $u^\mu$, as well as the shear-stress tensor $\pi^{\mu\nu}$ and the bulk-viscous pressure.\footnote{At high collision energies we can neglect the baryon chemical potential, i.e., we set $\mu_B=0$.} In this work we neglect viscous terms and only consider thermal Bose-Einstein and Fermi-Dirac distributions for primary particles on the freeze-out surface. 

For the case of thermal equilibrium, the vector distribution $g^\mu_b$ can be written in a Lorentz-invariant way as a sum of two scalar functions, $f_{1,b}^\text{eq}$ and $f_{2,b}^\text{eq}$, which are only functions of particle energy in the fluid rest frame $\bar{E}_{\textbf{p}} = |p^\mu u_\mu|$  and (implicitly) scalars like temperature, masses and branching ratios~\cite{Mazeliauskas:2018irt} 
\begin{equation}
g^\mu_b(x,p) = f^{\text{eq}}_{1,b}(\bar E_\p)\left(p^\mu-\bar E_\p u^\mu\right) + f^\text{eq}_{2,b}(\bar E_\p) \bar E_\p u^\mu. \label{eq:idealgmu}
\end{equation}
$f_{i,b}^\text{eq}$ correspond to irreducible SO(3) representations that in the fluid rest frame transform under rotations as a scalar and a vector. The isotropic decays do not mix these components and one can obtain the final irreducible components by repeated application of the two-body or three-body decay map. 

Taking into account the spectral widths, the transformation rule for 2-body decay between the parent and child components $f_{i,a}$ and $f_{i,b}$ is given by
\begin{align} 
  |\bar \p^b| f_{i,b}^\text{eq}&(\bar E_\p^b, m_b) = \int_{m_b+t_c}^\infty\hspace{-0.4cm} dm_a \int_{t_c}^{m_a-m_b} \hspace{-0.4cm} dm_c \rho_c(m_c)\rho_a(m_a) \nonumber\\
  &\times B \frac{\nu_a}{\nu_b} \frac{m_a^2}{m_b^2}\frac{1}{2}\int_{-1}^1 dw  |\bar \p^a| f_{i,a}^\text{eq}\left(E(w), m_a\right) A_i^\text{eq}(w).\label{eq:recursive}
\end{align}
Here $t_c$ is the lower mass threshold for the spectral function of particle $c$, i.e., the smallest combined mass of $c$'s decay products. The upper mass threshold for particle $c$ is given by $m_a-m_b$.
The range of possible values of $m_b$ is given by $m_b\geq t_b$. By construction, we have that $m_a \geq t_b+t_c \geq t_a$.

In \cref{eq:recursive} the weight functions for the two irreducible components are
\begin{equation}
A_1^\text{eq} = \frac{Q(w)}{p(w)},\qquad
A_2^\text{eq} = \frac{ E(w) |\bar \p| }{\bar E_\p p(w)}\,,
\end{equation}
where $E(w)$,  $Q(w)$ and $p(w)$ are defined as
\begin{subequations}
\begin{align}
E(w) &\equiv \frac{m_a E^a_{b|c}\bar E_\p}{m_b^2}-w \frac{m_a p^a_{b|c}|\bar \p|}{m_b^2},\\
Q(w)&=
\frac{m_a E^a_{b|c}|\bar \p|}{m_b^2}-w \frac{m_a p^a_{b|c}\bar E_\p}{m_b^2},\\
p(w)&=\sqrt{E(w)^2-m_a^2}
\end{align}
\end{subequations}
For 3-body decays the equivalent formula to \cref{eq:recursive} follows straightforwardly from \cref{eq:3body}.

In the new version of the package {\FastReso}, we have implemented two-body decays for any of the particles having a spectral function. For three-body decays, we allow at most one of the child particles to have a spectral width\footnote{We have checked that for the considered decay list we do not have decays with more than one child particle with a spectral function in a three-body decay.}. Additional mass integrals in \cref{eq:recursive} increase the computation cost of evaluating the decay feed-down, where, e.g., for intermediate resonances with finite spectral width, we need to loop over both their mass and momentum grids.
The multidimensional integrals are performed using the deterministic Cuhre cubature algorithm from the Cuba library, with a GSL Vegas Monte Carlo integration used as an independent cross-check. 
Our default decay list PDG2016~\cite{ParticleDataGroup:2016lqr,Alba:2017hhe,Alba:2017mqu} contains in total $2677$ two-body decays and $489$ three-body decays\footnote{We neglected 38 four-body decays included in the list.}. Of these, to compute decay feed-down to positively charged pions we need to perform $1522$ two-body decays and $284$ three-body decays.
Including the spectral functions for resonances with $\Gamma_i/M_i > 0.05$, the computation of the full decay chain for positively charged pions, kaons, and protons takes $\sim 30$ minutes using 20 threads. More details about the numerical implementation are provided in \cref{sec:numerical_details}.

It is important to stress that using the \FastReso{} procedure the decay components only need to be computed and tabulated once for a given temperature. Then the freeze-out integral can be performed using \cref{eq:freezeout_w_decays} for an arbitrary velocity profile. This significantly speeds up the computation of resonance feed-down, since decays for different fluid cells do not need to be re-computed.
Therefore, although the initial computation of resonance feed-down is computationally expensive, it can be included in event-by-event fluid dynamic simulations.

\section{Blast-wave spectra with resonance decay feed-down}
\label{sec:BlastWaveFits}

In order to estimate the impact of different resonance widths on the final particle spectra, we will use a parametrized freeze-out surface known as the blast-wave model~\cite{Schnedermann:1993ws}. In a boost-invariant and azimuthally symmetric blast-wave model, the freeze-out surface is given by $\tau=\tau_\text{fo}$ and $r<R$ with constant temperature $T_\text{fo}$. The transverse velocity is parametrized by a power-law function
\begin{equation}
\label{eq:betaT}
\beta_T = \dfrac{u^r}{u^{\tau}} = \beta_\text{max}\left( \dfrac{r}{R} \right)^{n}.
\end{equation}
The average transverse velocity $\langle\beta_{T}\rangle$ is then
\begin{equation}
\label{eq:avg_betaT}
    \langle \beta_{T}\rangle  = \frac{2}{2+n}\beta_\text{max}\,.
\end{equation}

The final particle momentum distribution produced from a fluid cell on this surface moving with a four-velocity $u^{\mu}$ can be estimated with \cref{eq:freezeout_w_decays}. In the case of an azimuthally symmetric and boost-invariant blast-wave surface, the Cooper-Frye integral reduces to a 1-dimensional integral~\cite{Mazeliauskas:2019ifr, Florkowski:2010zz}
\begin{equation}
\label{eq:blast_wave_spec}
\dfrac{dN}{2\pi p_T dp_T dy} =  \dfrac{\nu}{\left( 2\pi\right)^{3}} \int_0^R dr\, \tau_\text{fo} r K_1^\text{eq}(p_T, \beta_T(r)).
\end{equation}

Here, the freeze-out kernel $K_1^\text{eq}(p_T, \beta_T)$ is defined as an azimuthal and space-time rapidity integral over the scalar distributions $f_{i = 1, 2}^\text{eq}$ which contain the decay feed-down of unstable hadrons
\begin{equation}
\label{eq:CF_kernels}
\begin{aligned}
K_1^\text{eq}(p_T, \beta_T) &= \int_0^{2\pi}d\phi \int_{-\infty}^{\infty} d\eta \lbrace f_1^\text{eq}\left( \bar{E}_\textbf{p}\right) m_T \cosh \left(\eta\right)\\
			& + \left( f_2^\text{eq}\left(\bar{E}_\textbf{p}\right) - f_1^\text{eq}\left(\bar{E}_\textbf{p}\right) \right)\bar E_\textbf{p} u^{\tau} \rbrace\, .
\end{aligned}
\end{equation}
 Here, $\bar{E}_\textbf{p}$ takes the form $\bar{E}_\textbf{p}= m_T u^{\tau} \cosh (\eta) - p_T u^{r} \cos \phi$, where $m_T = \sqrt{p_T^{2} + M^2}$ is the transverse mass. The radial four-velocity is given by $u^{r} = \beta_T/\sqrt{1-\beta_T^2}$ and $u^\tau =1/\sqrt{1-\beta_T^2}$.

The decay kernel $K_1^\text{eq}$ can be precomputed for a specified range of transverse momentum $p_T$, velocity $\beta_T$, and freeze-out temperature $T_\text{fo}$. Then, finding the particle spectra can be done by a one-dimensional integral in \cref{eq:blast_wave_spec}. This allows a systematic variation of model parameters, such as $\beta_\text{max}$, $n$, $R$, $T_\text{fo}$, to obtain the best fit to the measured particle spectra~\cite{Mazeliauskas:2019ifr}. 
 We assume that all primary particles are produced along the same freeze-out surface, therefore their spectra are proportional to the freeze-out volume per rapidity (in the lab frame) $dV/dy = \tau_\text{fo} \pi R^2$. Since in this model the radial size, $R$, and the time $\tau_\text{fo}$ appear only in this combination, we cannot constrain them separately and we just set $\tau_\text{fo}=R$. 
The parameters $\left( \beta_\text{max}, T_\text{fo}, R, n\right)$ of this simple model are extracted by simultaneously fitting the experimentally measured transverse momentum spectra of charged pions ($\pi$), charged kaons ($K$) and protons/antiprotons ($p$)\footnote{We computed decay feed-down for positively charged particles. The negative partners have identical distributions in the $\mu_B=0$ case.}.

In this work we compute the decay kernels of $\pi$, $K$ and $p$ using \cref{eq:CF_kernels} for the PDG2016 particle list~\cite{ParticleDataGroup:2016lqr,Alba:2017hhe,Alba:2017mqu} and four cases of increasing sophistication of particle spectral functions.
\begin{description}
    \item[Dirac] all particles have a vanishing spectral width, i.e., a Dirac-$\delta$-like spectral function at the pole mass.
    \item[Breit-Wigner] all particles satisfying $\Gamma_i/M_i>0.05$ are described by a Breit-Wigner distribution, \cref{eq:breitwigner}. This represents more than 80\% of all particles listed in PDG2016.\footnote{A cutoff of $\Gamma_i/M_i>0.1$ would instead cover just over 50\% of the particles listed in PDG2016. Tightening the cutoff from $0.1$ to $0.05$ changes the final $\pi$-spectrum by less than 3\%, indicating that the results are not sensitive to the precise choice of cutoff.} For particles with $\Gamma_i/M_i<0.05$ the spectral width is neglected as in the Dirac case.
    \item[S-matrix] $\rho$-mesons and $\Delta$-baryons are parametrized using the S-matrix formulation of the spectral function, see \cref{sec:smatrix}. Other particles are described as in the Breit-Wigner case.
    \item[In-medium] $\rho$, $\omega$, and $\phi$ are parametrized with in-medium spectral functions, see \cref{sec:inmedium}, $\Delta$-baryons are parametrized using S-matrix formulation and all other particles are described as in the Breit-Wigner case.
\end{description}

The kernels are precomputed in the ranges $p_T\in[0,4]\,\si{\GeV}$, $\beta_T\in[0.0,0.96]$ and temperature range $T\in[130,155]\,\si{\MeV}$ with $\Delta T=\SI{5}{\MeV}$ steps.

\subsection{Comparisons at constant temperature}

\begin{figure}
\centering\includegraphics[width=1\linewidth]{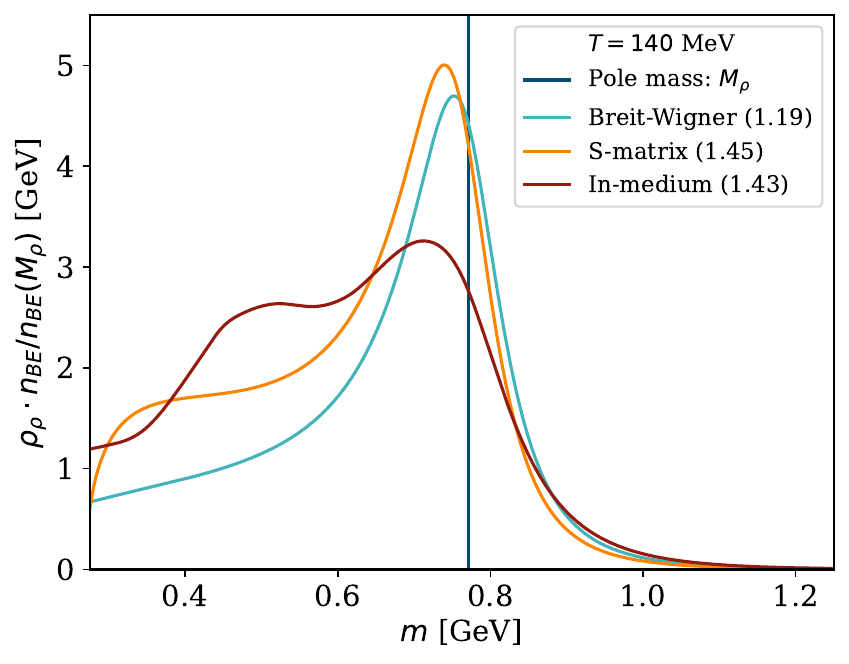}
    \caption{Spectral function of the $\rho$ meson weighted by a thermal density $n_\text{BE}(m)$, i.e., momentum-integrated Bose-Einstein distribution for mass $m$ and $T = \SI{140}{\MeV}$.
        In the legend, we report the mass-integrated thermal $\rho$ number ratio to the Dirac case, i.e., density at the pole mass.}
    \label{fig:spectralfunction_times_nBE}
\end{figure}

We first illustrate how a finite width modifies the thermal abundance of a broad
resonance itself, before any decay is performed.
\Cref{fig:spectralfunction_times_nBE} shows the mass distribution of thermally
populated $\rho$ mesons at $T=\SI{140}{\MeV}$, i.e., the $\rho$ spectral
function weighted by the momentum-integrated Bose-Einstein density $n_\text{BE}(m)$.
Two effects compete here: the spectral function suppresses states far below the pole
mass, while the thermal occupancy grows exponentially, $n_\text{BE}\sim e^{-m/T}$, as
the mass decreases. Integrated
over mass, the thermal $\rho$ number exceeds its Dirac (zero-width) value by $19\%$ for
the Breit-Wigner parametrization and by more than $40\%$ for the S-matrix and In-medium
spectral functions, as listed in the legend of
\cref{fig:spectralfunction_times_nBE}. We note that the S-matrix spectral function is
constructed to vanish at the mass threshold, whereas the Breit-Wigner and In-medium
parametrizations remain finite there.

The enhancement becomes more dramatic for heavy resonances whose decay thresholds lie
far below their pole mass, to the point where the thermal weighting can move the maximum
of the mass distribution below the pole. Such states, however, are not free to decay.
The child mass integral in \cref{eq:recursive} is bounded from above by
$m_c \le m_a - m_b$, so a parent populated below its pole mass can only access the
low-mass tail of a broad child's spectral function and the channel is strongly suppressed. Because the branching ratios $B$ are
measured at the pole mass, the channel probabilities of such a parent no longer add up
to unity --- formally, not all thermally initialized resonances decay. We keep the constant
pole-mass branching ratios throughout. A consistent
treatment, for instance through mass-dependent partial widths or a renormalization of
the branching ratios to the accessible phase space, is left for future work.

 \begin{figure*}
    \includegraphics[width=1.0\linewidth]{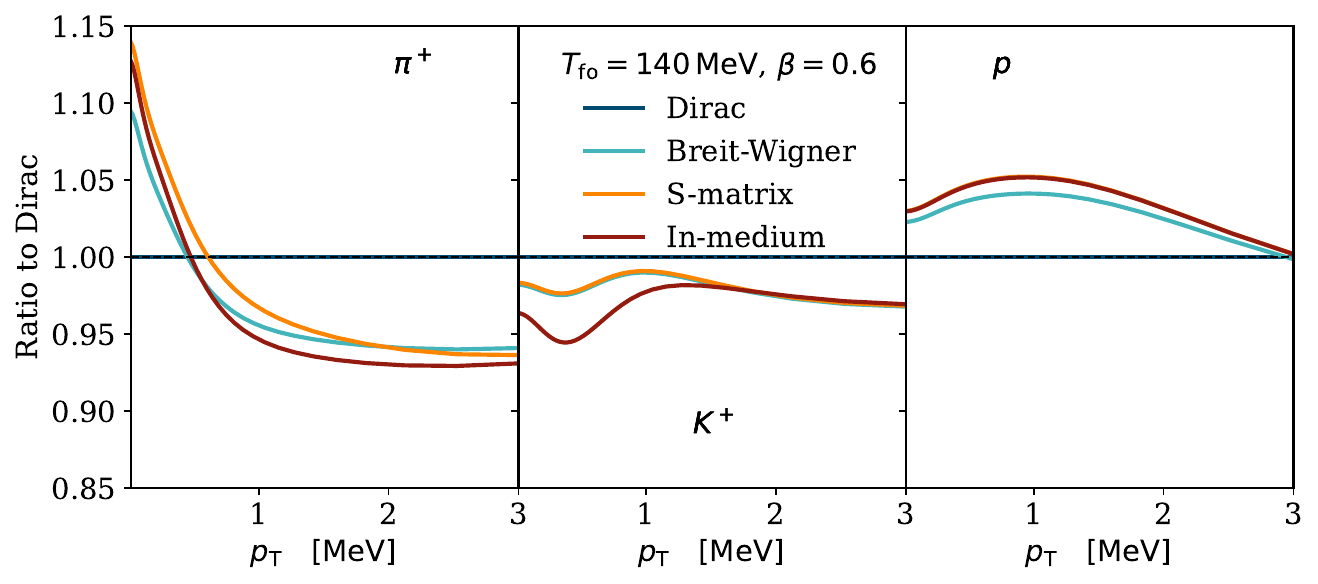}
    \caption{Ratio of $\pi$ (left), $K$ (middle) and $p$ (right) spectra (after resonance decays) at constant fluid velocity $\beta_T=0.6$ and temperature $T_\text{fo}=\SI{140}{\MeV}$ for different spectral functions. The Dirac spectrum is used as a reference.}
    \label{fig:Kj_ratios}
\end{figure*}

We now turn to the effect of the spectral widths on the full resonance decay chain. We consider the resonance decay feed-down from a constant velocity $\beta_T = 0.6$ and temperature $T = \SI{140}{\MeV}$ freeze-out surface. In this unrealistic, but simple, scenario, \cref{eq:blast_wave_spec} reduces to the decay kernel $K_1^\text{eq}(p_T, \beta_T)$ times the volume element. Taking the
ratio to the Dirac case, the volume element cancels and the results for
$\pi$, $K$ and $p$ are shown in \cref{fig:Kj_ratios}. 

For pions, \cref{fig:Kj_ratios}~(left), the Breit-Wigner spectral functions increase the
spectra by up to $10\%$ at $p_T<\SI{0.5}{\GeV}$ relative to the Dirac case, but suppress
them by $\sim 5\%$ for $p_T>\SI{1}{\GeV}$, before turning upward towards hard modes, $p_\mathrm{T}\sim\SI{3}{\GeV}$. This pattern follows the shape of the $\rho$
spectral function in \cref{fig:spectralfunction_times_nBE}: the spectral weight removed
from the pole region suppresses pion production at intermediate momenta, while the
redistributed strength enhances both the soft and hard pions.

Since the overall normalization is a free parameter when the blast-wave spectra are
refitted to data, the relevant quantity is the change in the shape of the
spectrum. Measured against $p_T\sim \SI{1}{\GeV}$, the Breit-Wigner case enhances the
lowest-momentum pions by $\sim 15\%$ with respect to the Dirac case. Note that the decay kernels include the thermal pion contribution from the freeze-out spectra, which at $T_\mathrm{fo} = \SI{140}{\MeV}$ constitutes around 35--55\% of pions with momenta below \SI{1}{\GeV}\footnote{The contribution coming from primordial particles to the kernels, and thus to the final spectra, depends on the freeze-out temperature, considered momentum range and is species-dependent, see \cref{sec:ThermalFraction}.}. Therefore, the modification of
the decay contribution alone is correspondingly larger. 

Using the S-matrix parametrization enhances the soft-pion
spectra by up to $\sim 14\%$ and pushes the onset of the high-momentum suppression out to
$p_T\sim \SI{2}{\GeV}$. The In-medium spectral functions do not increase the soft-pion
enhancement any further, but deepen the suppression at $p_T>\SI{1}{\GeV}$
slightly. Of the three choices, the In-medium spectral functions
therefore produce the largest relative enhancement of low- over high-momentum pions.

For kaons, \cref{fig:Kj_ratios}~(middle), the Breit-Wigner and S-matrix scenarios coincide
and give an overall suppression of about $2\%$ over most of the displayed $p_T$ range.
The two agree because the S-matrix parametrization is applied only to the $\rho$ and
$\Delta$, neither of which feeds the kaons. Both show a minimum at
$p_T \simeq \SI{0.5}{\GeV}$, which becomes more pronounced for the In-medium spectral
functions. This is likely a consequence of the melting of the $\phi$ meson in the in/medium,
since the $\phi$ is the dominant broad kaon parent. 

Finally, for protons, \cref{fig:Kj_ratios}~(right), all spectral functions produce a broad
net enhancement. The Breit-Wigner case gives an increase of up to $4\%$, which grows to
$\sim 6\%$ around $p_T \simeq \SI{1}{\GeV}$ for
S-matrix $\Delta$ spectral function. The In-medium curve overlaps with the
S-matrix one, as expected: the In-medium scenario differs from it only in the $\rho$,
$\omega$ and $\phi$ spectral functions, which do not contribute to the proton feed-down.

Throughout \cref{fig:Kj_ratios} the freeze-out temperature, velocity and volume were held
fixed across all four scenarios. A more realistic assessment requires refitting the
blast-wave parameters separately for each choice of spectral function, which we do in the
following section.

\section{Results}
\label{sec:Results}

\begin{figure*}
    \centering
    \includegraphics[width=0.33\linewidth]{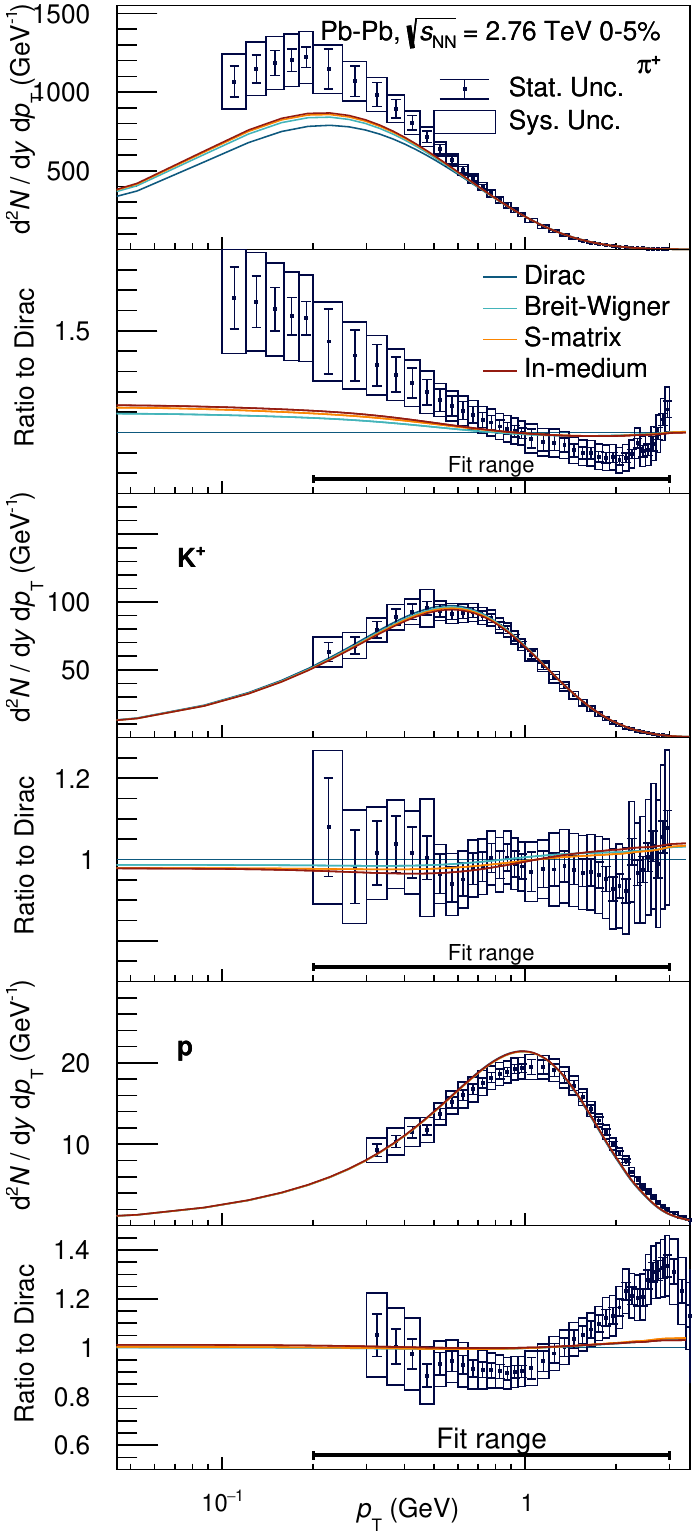}%
    \includegraphics[width=0.33\linewidth]{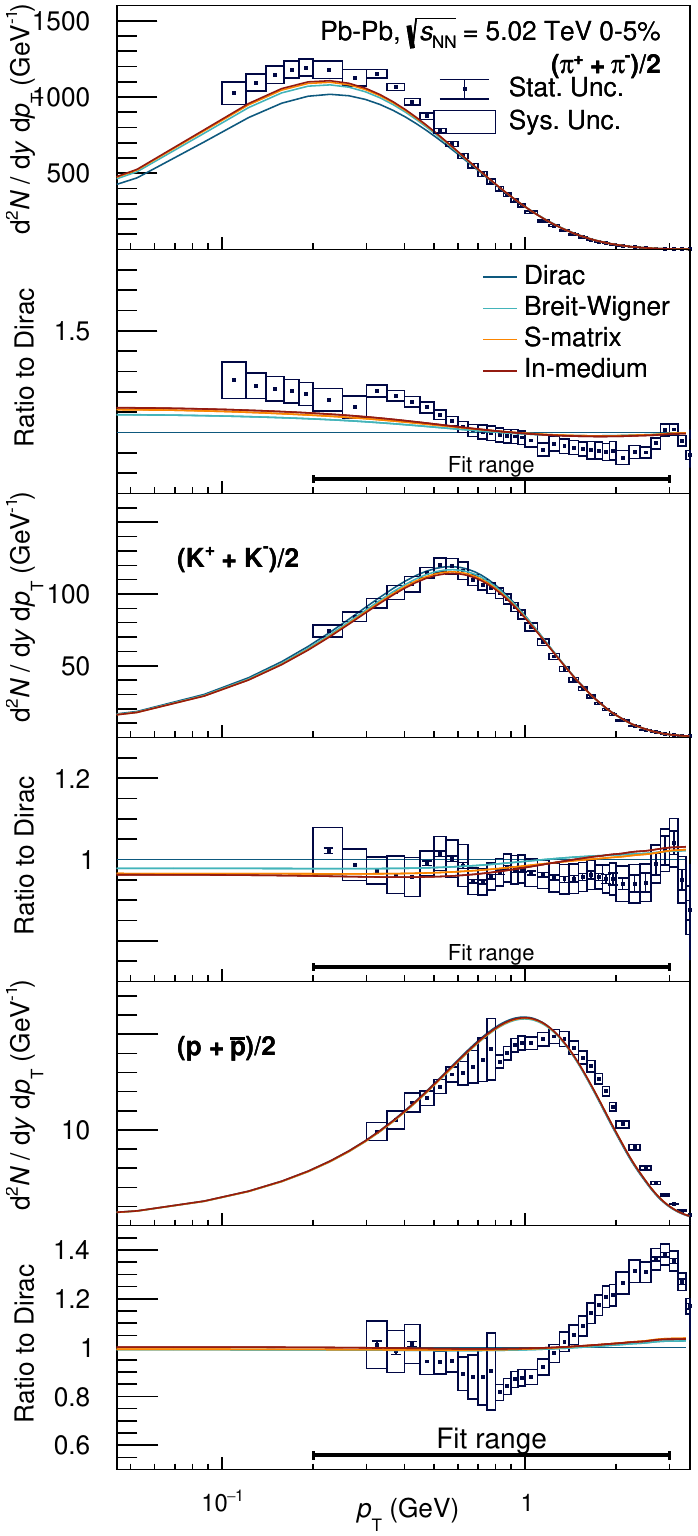}%
    \includegraphics[width=0.33\linewidth]{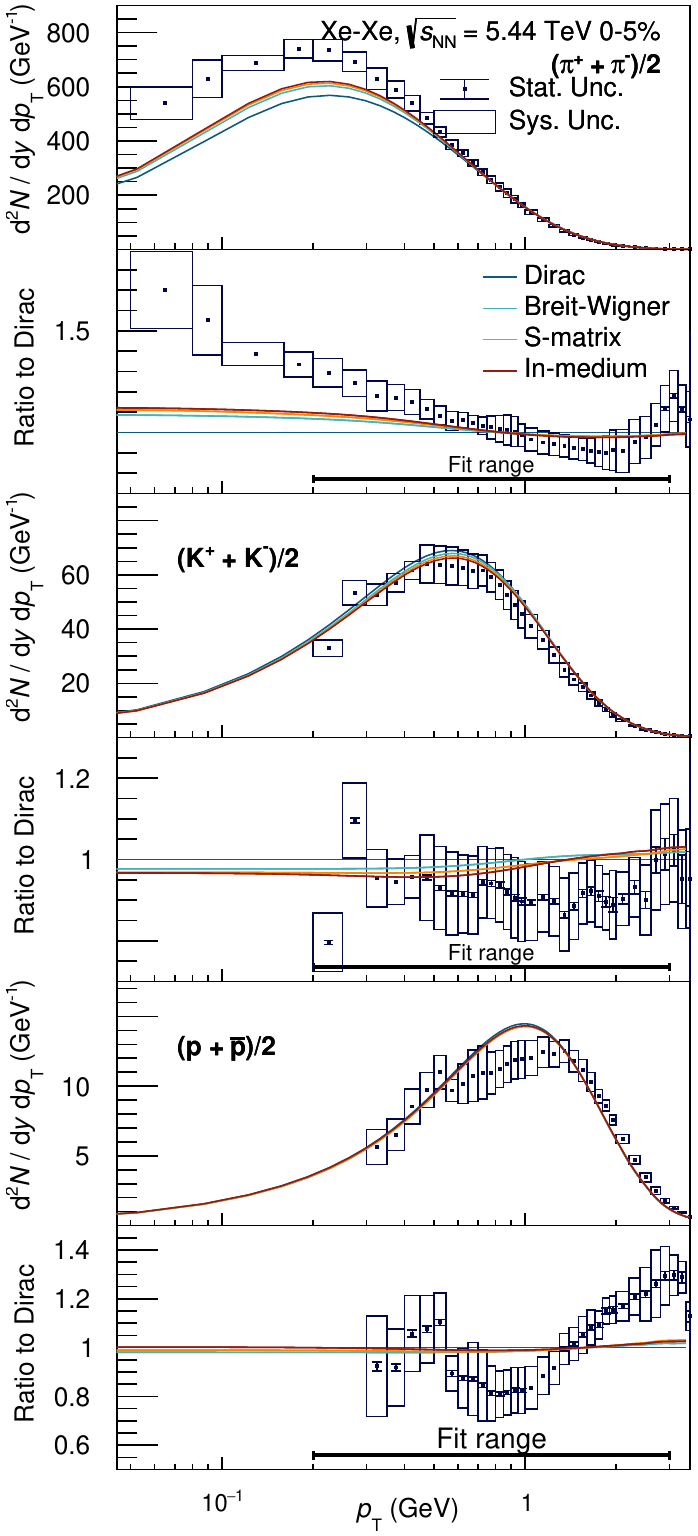}
    \caption{Comparison of the fitted blast-wave model with resonance feed-down to the experimental data for $\pi^+$ (top row), $K^+$ (middle row) and $p$ (bottom row) for the four spectral function scenarios. The columns show Pb-Pb collisions at $\sqrt{s_{NN}} = \SI{2.76}{\TeV}$ (left) and $\SI{5.02}{\TeV}$ (middle) and Xe-Xe collisions at $\sqrt{s_{NN}} = \SI{5.44}{\TeV}$ (right). In each panel the upper part shows the particle spectrum and the lower part the ratio of the fits and of the measured data to the Dirac fit, i.e., to the fit assuming vanishing spectral widths for all particles.}
    \label{fig:spectra_fits}
\end{figure*}

In this section we present the results of blast-wave fits including resonance decay feed-down to the charged pion, kaon and proton spectra measured by the ALICE Collaboration in $0$--$5\%$ central Pb-Pb collisions at $\sqrt{s_{NN}} = \SI{2.76}{\TeV}$~\cite{ALICE:2013mez} and $\SI{5.02}{\TeV}$~\cite{ALICE:2019hno}, and in Xe-Xe collisions at $\sqrt{s_{NN}} = \SI{5.44}{\TeV}$~\cite{ALICE:2021lsv}. For each system we compare the four spectral function scenarios (Dirac, Breit-Wigner, S-matrix and In-medium) introduced in \cref{sec:BlastWaveFits}.
Our goal is to quantify the sensitivity of light hadron spectra to the choice of spectral function and to test whether in-medium broadening can account for the excess of soft pions seen in the data, which is not reproduced by full hydrodynamic simulations~\cite{Devetak:2019lsk,Nijs:2020ors,Lu:2024shm}.
Since our model uses a simplified freeze-out surface and neglects viscous corrections and hadronic rescattering, we focus on the relative differences between the spectral function scenarios.

We fit the transverse momentum spectra in the range $\SI{0.2}{\GeV} \leq p_T \leq \SI{3.0}{\GeV}$, which is somewhat wider than the ranges used in previous works, see Refs.~\cite{Mazeliauskas:2019ifr,Melo:2019mpn,Lu:2024shm}. We perform a least-squares fit using the \textsc{Minuit2} package of \textsc{ROOT}~\cite{James:1975dr,Hatlo:2005ts},
i.e., we minimize
\begin{align}
\chi^2 &= \chi^2_\pi +\chi^2_K+\chi^2_p,\\
    \chi^2_a &= \sum_{i} \frac{\left(N_a^\text{exp}(p_T^i) - N_a^\text{model}(p_T^i)\right)^2}{\sigma^2_{a,i}}\,.
\end{align}
with respect to the model parameters $\beta_\text{max}$, $n$, $R$ and $T_\text{fo}$. Here $a=\pi,K,p$ labels the particle species, $N_a(p_T)\equiv dN_a/(2\pi p_T dp_T dy)$ is the spectrum of \cref{eq:blast_wave_spec} averaged over the $i$-th measured momentum bin $p_T^i$, and $\sigma_{a,i}$ is the corresponding experimental statistical and systematic uncertainty added in quadrature. The number of degrees of freedom is the total number of data points minus the four fit parameters. We neglect the correlations of systematic uncertainties, which might result in the underestimation of $\chi^2$.

\begin{table*}
\begin{center}
 \renewcommand{\arraystretch}{1.3}
\begin{tabular}{|l||c|c|c|c||c|c|c|c|}
\hline
 & \multicolumn{4}{c||}{Pb-Pb \SI{2.76}{\TeV}} & \multicolumn{4}{c|}{Pb-Pb \SI{5.02}{\TeV}} \\
 \cline{2-9}
 & \multicolumn{4}{c||}{Spectral Function} & \multicolumn{4}{c|}{Spectral Function} \\
 & Dirac & Breit-Wigner & S-matrix & In-medium & Dirac & Breit-Wigner & S-matrix & In-medium \\
\hline
$T_\text{fo}$ [MeV] &
$148.0\pm0.7$ &
$145.8\pm0.5$ &
$145.9\pm0.6$ &
$145.2\pm0.5$ &
$141.1\pm0.4$ &
$139.2\pm0.4$ &
$139.6\pm0.5$ &
$138.8\pm0.4$ \\
$\langle\beta_T\rangle$ &
$0.638\pm0.010$ &
$0.639\pm0.009$ &
$0.639\pm0.009$ &
$0.638\pm0.009$ &
$0.638\pm0.005$ &
$0.640\pm0.005$ &
$0.640\pm0.005$ &
$0.639\pm0.005$ \\
$dV/dy$ [fm$^3$] &
$4732\pm191$ &
$5388\pm182$ &
$5313\pm180$ &
$5626\pm180$ &
$8296\pm174$ &
$9148\pm204$ &
$8889\pm200$ &
$9359\pm201$ \\

$n$ &
$0.46\pm0.03$ &
$0.49\pm0.03$ &
$0.49\pm0.03$ &
$0.50\pm0.03$ &
$0.608\pm0.018$ &
$0.622\pm0.018$ &
$0.622\pm0.018$ &
$0.626\pm0.018$ \\
$\chi^2/\mathrm{dof}$ &
$2.2$ &
$1.8$ &
$1.6$ &
$1.6$ &
$6.5$ &
$5.2$ &
$4.8$ &
$4.8$ \\
$\chi^2_{\pi^{+}}/N_{\pi^{+}}$ &
$3.4$ &
$2.6$ &
$2.3$ &
$2.1$ &
$6.9$ &
$4.2$ &
$3.7$ &
$3.3$ \\
$\chi^2_{K^{+}}/N_{K^{+}}$ &
$0.17$ &
$0.23$ &
$0.19$ &
$0.24$ &
$0.81$ &
$0.84$ &
$0.66$ &
$0.80$ \\

$\chi^2_{p}/N_{p}$ &
$2.8$ &
$2.3$ &
$2.2$ &
$2.4$ &
$11$ &
$10$ &
$9.7$ &
$9.9$ \\
\hline
\end{tabular}

  \renewcommand{\arraystretch}{1}
 \end{center}

\begin{center}
 \renewcommand{\arraystretch}{1.3}

\begin{tabular}{|l||c|c|c|c|}
\hline
 & \multicolumn{4}{c|}{Xe-Xe \SI{5.44}{\TeV}} \\
 \cline{2-5}
 & \multicolumn{4}{c|}{Spectral Function} \\
 & Dirac & Breit-Wigner & S-matrix & In-medium \\
\hline
$T_\text{fo}$ [MeV] & $146.4\pm0.7$ & $144.0\pm0.8$ & $144.3\pm0.8$ & $143.5\pm0.7$ \\
$\langle\beta_T\rangle$ & $0.641\pm0.010$ & $0.643\pm0.009$ & $0.643\pm0.009$ & $0.642\pm0.009$ \\
$dV/dy$ [fm$^3$] & $3685\pm136$ & $4204\pm157$ & $4089\pm154$ & $4319\pm154$ \\
$n$ & $0.53\pm0.04$ & $0.55\pm0.03$ & $0.55\pm0.03$ & $0.56\pm0.03$ \\
$\chi^2/\mathrm{dof}$ & $2.3$ & $1.8$ & $1.6$ & $1.6$ \\
$\chi^2_{\pi^{+}}/N_{\pi^{+}}$ & $2.7$ & $1.6$ & $1.4$ & $1.2$ \\
$\chi^2_{K^{+}}/N_{K^{+}}$ & $1.2$ & $1.2$ & $1.0$ & $1.1$ \\
$\chi^2_{p}/N_{p}$ & $2.7$ & $2.4$ & $2.3$ & $2.4$ \\
\hline
\end{tabular}%

\end{center}
\caption{Parameters extracted from simultaneous blast-wave fits to the $\pi^+$, $K^+$ and $p$ spectra in $p_T\in[0.2,3.0]\,\si{\GeV}$ measured in $0$--$5\%$ central Pb-Pb collisions at $\sqrt{s_{NN}}=\SI{2.76}{\TeV}$ and $\SI{5.02}{\TeV}$ (top) and Xe-Xe collisions at $\sqrt{s_{NN}}=\SI{5.44}{\TeV}$ (bottom), for the four spectral function scenarios and the PDG2016 decay list, as shown in \cref{fig:spectra_fits}. The average transverse velocity $\langle\beta_T\rangle$ is derived from the fit parameters $\beta_\text{max}$ and $n$ using \cref{eq:avg_betaT}. The last four rows give the reduced $\chi^2$ of the combined fit and the $\chi^2_a$ of each species normalized by its number of data points $N_a$.}
  \label{tab:fitparams}
\end{table*}

In \cref{fig:spectra_fits} we show the results of our fits to the final spectra of $\pi^+$, $K^+$ and $p$ or $(\pi^++\pi^-)/2$, $(K^++K^-)/2$ and $(p+\bar p)/2$, with the corresponding fit parameters summarized in \cref{tab:fitparams}.
The left column corresponds to Pb-Pb at $\sqrt{s_{NN}} = \SI{2.76}{\TeV}$, the middle column to Pb-Pb at $\sqrt{s_{NN}} = \SI{5.02}{\TeV}$ and the right column to Xe-Xe at $\sqrt{s_{NN}} = \SI{5.44}{\TeV}$.
The top row shows the results for pions, the middle row for kaons and the bottom row for protons.
For each particle species the plot is divided into two panels: the upper panel compares the measured spectra with the fits for the four scenarios (Dirac, Breit-Wigner, S-matrix and In-medium), while the lower panel shows the same curves normalized to the Dirac fit. The fit range is indicated in the plots.

For pions in the top row of \cref{fig:spectra_fits} we see that all fits fail to describe the low-momentum region $p_T< \SI{0.5}{\GeV}$. The discrepancy is largest for the Dirac case, reaching up to 65\% for Pb-Pb at \SI{2.76}{\TeV}, 25\% for Pb-Pb at \SI{5.02}{\TeV} and 70\% for Xe-Xe at \SI{5.44}{\TeV}, admittedly with sizable systematic experimental uncertainties. Introducing a finite spectral width increases the low-momentum pion spectra and reduces this discrepancy. For Breit-Wigner spectral functions the refitted pion spectrum is about 10\% higher than in the Dirac case, with a corresponding reduction of the discrepancy.
The S-matrix and In-medium spectral functions give a slightly larger soft pion yield, the In-medium case being the largest, although the difference between the two is rather small. These results are consistent with the trend observed in the decay kernels shown in \cref{fig:Kj_ratios}~(left).
We stress, however, that the fit parameters differ between the four scenarios, as summarized in \cref{tab:fitparams}, so that the comparison in \cref{fig:spectra_fits} is made at the best fit of each scenario rather than at fixed freeze-out parameters.

In the middle row of \cref{fig:spectra_fits} we show the results for kaons. The approximately momentum-independent decrease of the kaon yield (within 5\%) caused by the spectral functions, seen in \cref{fig:Kj_ratios}, is compensated by the change of the freeze-out parameters in the refit, so that the four scenarios differ by only a few percent. We note, however, that at low momenta the kaon fits with finite widths lie below the Dirac case, i.e., the trend is opposite to the one observed for pions, but expected from the intuition obtained from \cref{fig:Kj_ratios}.
For Pb-Pb at \SI{2.76}{\TeV} (left column) the kaon spectra are very well described in all four scenarios throughout the fitted momentum range. 
For Pb-Pb at \SI{5.02}{\TeV} (middle column) and for Xe-Xe at \SI{5.44}{\TeV} (right column) we see a slight overestimation of the kaon spectrum in the $1$--$\SI{2}{\GeV}$ range.

In the bottom row of \cref{fig:spectra_fits} we show the results for protons. After refitting, the blast-wave proton spectrum barely changes between the different spectral function scenarios. For Pb-Pb at \SI{2.76}{\TeV} the proton spectrum is reasonably well described below \SI{2}{\GeV}, although the ratio to the data shows a small dip around \SI{1}{\GeV} and rises sharply at larger momenta. For Pb-Pb at \SI{5.02}{\TeV} and Xe-Xe at \SI{5.44}{\TeV} this trend is more pronounced. As seen in the absolute spectra, the proton spectra are shifted to lower momenta compared to the data.

\Cref{tab:fitparams} summarizes the best-fit parameters for the three fitted systems together with the breakdown of $\chi^2$ by particle species.
Including finite spectral widths lowers the freeze-out temperature by $\sim1.5$--$\SI{3}{\MeV}$ in all three systems and increases the freeze-out volume $dV/dy$ by $\sim 10-20\%$. Both shifts are several times larger than the quoted fit uncertainties, whereas the differences between the three finite-width scenarios are comparable to them.
The average transverse velocity $\langle\beta_T\rangle$ and the velocity exponent $n$, in contrast, are unchanged within uncertainties for all scenarios.
These trends follow from the effect of the widths on the decay kernels seen in \cref{fig:Kj_ratios}: the proton spectrum is slightly enhanced, while pions above \SI{1}{\GeV} and kaons across the entire fit range are slightly suppressed. To recover the measured spectra, the fit therefore prefers a lower freeze-out temperature, which increases the relative fraction of pions to kaons, together with a larger freeze-out volume $dV/dy=\pi \tau_\text{fo} R^2$, which compensates for the overall decrease of the particle density at freeze-out.
Comparing the three systems, the freeze-out temperature decreases and the volume grows from \SI{2.76}{\TeV} to \SI{5.02}{\TeV} Pb-Pb collisions, while the smaller Xe-Xe system at \SI{5.44}{\TeV} freezes out at a temperature close to that of the lower-energy Pb-Pb collisions. These orderings are the same for all four spectral function scenarios.

Turning to the fit quality, the inclusion of spectral functions systematically improves the agreement between the blast-wave model and the data: $\chi^2/\mathrm{dof}$ decreases from $2.2$ to $1.6$ for Pb-Pb at \SI{2.76}{\TeV}, from $6.5$ to $4.8$ for Pb-Pb at \SI{5.02}{\TeV} and from $2.3$ to $1.6$ for Xe-Xe at \SI{5.44}{\TeV}, going from the Dirac to the In-medium case.
The improvement is driven almost entirely by pions, whose $\chi^2_{\pi^+}/N_{\pi^+}$ drops from $3.4$ to $2.1$, from $6.9$ to $3.3$ and from $2.7$ to $1.2$ in the three systems, the In-medium scenario giving the lowest value in each case.
For kaons, $\chi^2_{K^+}/N_{K^+}$ stays close to or below unity in all three systems and is largely insensitive to the choice of spectral function, consistent with the almost momentum-independent suppression of the soft kaon spectra discussed above.
For protons, $\chi^2_p/N_p$ is also largely insensitive to the choice of spectral function, indicating that changing it does not improve the fit. The
mismatch in the momentum trend around $\SI{2}{\GeV}$ is largely responsible for the large $\chi^2$, particularly for Pb-Pb at \SI{5.02}{\TeV}, where it is around $10$.
We recall that we treat the sizable experimental systematic uncertainties as uncorrelated, which can underestimate $\chi^2$ --- most visibly for kaons --- and therefore overstate the quality of the description.

\begin{figure}
\centering
\includegraphics[width=\linewidth]{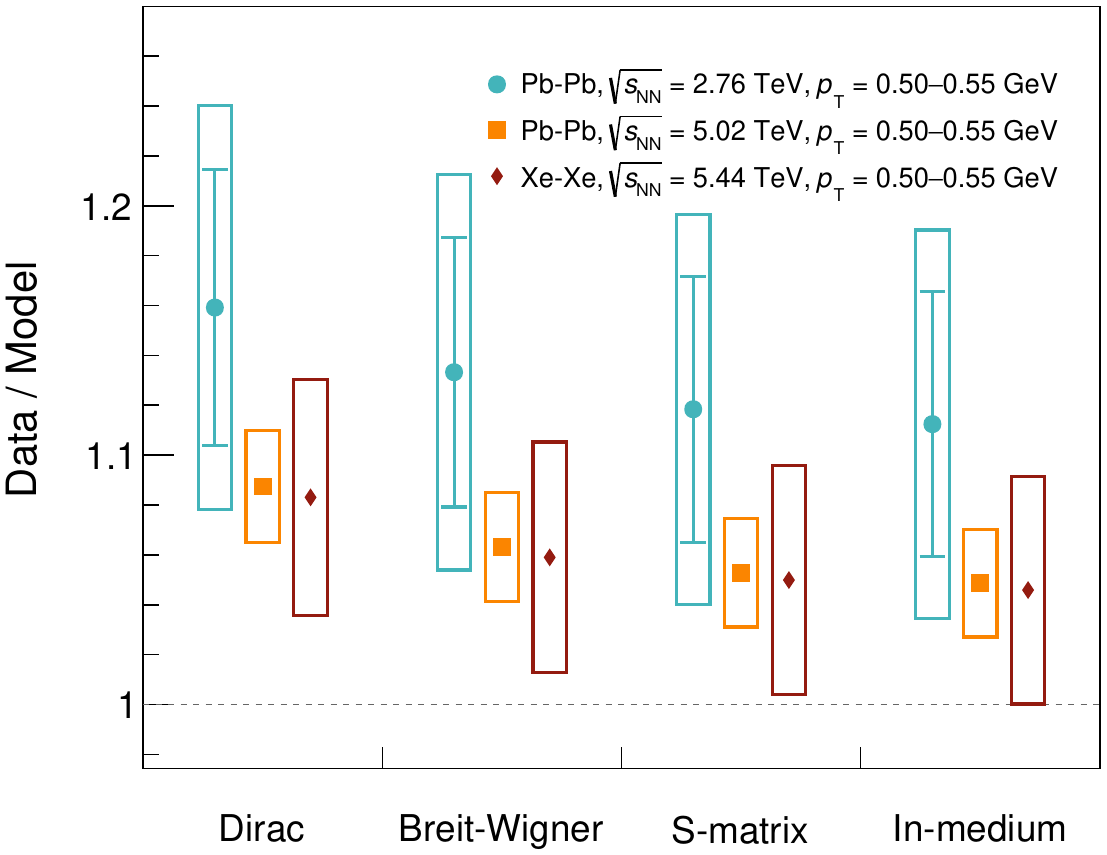}\\
\includegraphics[width=\linewidth]{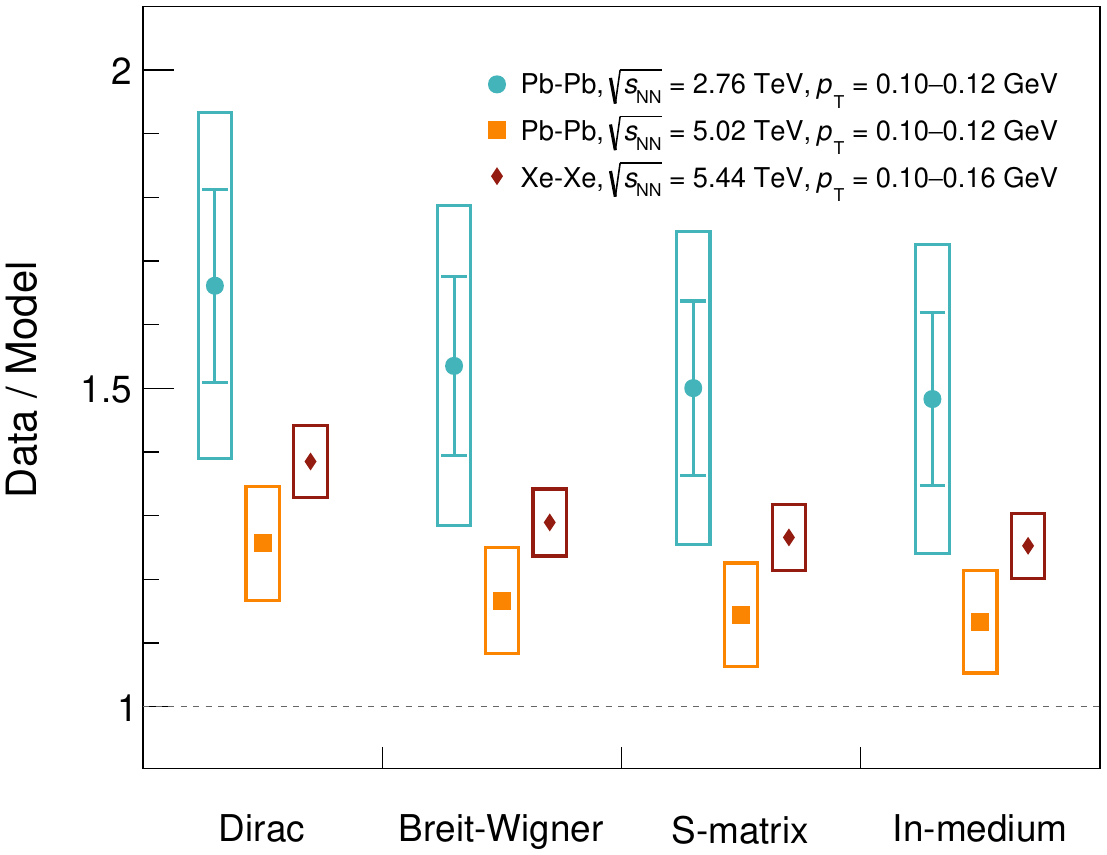}
\caption{Ratio of the measured $\pi^+$ or $(\pi^++\pi^-)/2$ yield to the fitted blast-wave model with resonance feed-down, for the four spectral function scenarios and the three collision systems of \cref{fig:spectra_fits}. The top panel corresponds to $p_T=0.50$--$\SI{0.55}{\GeV}$ and the bottom panel to $p_T=0.10$--$\SI{0.12}{\GeV}$ ($0.10$--$\SI{0.16}{\GeV}$ for Xe-Xe). Boxes indicate systematic and bars statistical uncertainties.}
\label{fig:DatatoModel_comp}
\end{figure}

Finally, in \cref{fig:DatatoModel_comp} we compare the pion enhancement in two reference momentum bins, $p_T\sim \SI{0.5}{\GeV}$ (top) and $p_T\sim \SI{0.1}{\GeV}$ (bottom), between the different spectral function scenarios and the three collision systems. For $p_T=0.50$--$\SI{0.55}{\GeV}$ the data-to-model discrepancy decreases with increasing sophistication of the spectral widths, from the Dirac to the In-medium case. The three systems show a consistent discrepancy, with Pb-Pb at \SI{5.02}{\TeV} and Xe-Xe at \SI{5.44}{\TeV} giving almost identical central values, albeit with large systematic uncertainties. The Pb-Pb points at \SI{2.76}{\TeV} lie systematically above the other two, but they also carry large statistical and systematic uncertainties.
The lower panel shows the results for $p_T=0.10$--$\SI{0.12}{\GeV}$ in Pb-Pb and $p_T=0.10$--$\SI{0.16}{\GeV}$ in Xe-Xe. Here Pb-Pb at \SI{5.02}{\TeV} lies systematically below the other two systems, reflecting the discontinuous shift of the measured spectrum in the lowest momentum bins that is visible in \cref{fig:spectra_fits}. 
The data-to-model ratio decreases for the In-medium spectral functions, but remains above unity: $1.5\pm0.3$ for Pb-Pb at $\SI{2.76}{\TeV}$, $1.13\pm0.08$ for Pb-Pb at $\SI{5.02}{\TeV}$ and $1.25\pm0.05$ for Xe-Xe at $\SI{5.44}{\TeV}$.
 In summary, the introduction of in-medium spectral functions enhances the production of low-momentum pions, but is not sufficient to bring the data-to-model ratio to unity.

\section{Conclusions}
\label{sec:conclusions}
In the present paper, we have studied the sensitivity of the soft-pion spectrum measured in heavy-ion collisions to the parametrization of the spectral function of broad particles, i.e., particles fulfilling $\Gamma_i/M_i > 0.05$, at the freeze-out surface.
We considered vacuum Breit-Wigner spectral functions and the S-matrix parametrization of Refs.~\cite{Dashen:1969ep,Huovinen:2016xxq,Lo:2017ldt} for the $\rho$ meson and the $\Delta$ baryon, and improved on previous analyses by introducing in-medium spectral functions for the light mesons relevant for the light hadron spectrum ($\rho,\, \omega,$ and $\phi$).
Since hadrons in heavy-ion collisions freeze out close to thermal equilibrium at a temperature close to the QCD chiral crossover, the thermal broadening
of light vector mesons is a genuine and previously unaccounted-for contribution to the soft-pion excess.

To compute resonance decays with arbitrary spectral functions we extended the efficient resonance decay package {\FastReso}, which previously could only compute the feed-down of particles on-shell. With this extension we computed the irreducible decay functions, and hence the full resonance feed-down of two-body and three-body decays
from the PDG2016 particle list~\cite{ParticleDataGroup:2016lqr,Alba:2017hhe,Alba:2017mqu} with different levels of spectral function sophistication (see \cref{sec:BlastWaveFits}).
We have then performed simultaneous blast-wave fits (with resonance feed-down included) to $\pi^+$, $K^+$ and $p$ spectra measured by ALICE in $0$--$5\%$ central Pb-Pb collisions at $\sqrt{s_{NN}}=\SI{2.76}{\TeV}$ and $\SI{5.02}{\TeV}$, and Xe-Xe collisions at $\sqrt{s_{NN}}=\SI{5.44}{\TeV}$.

We showed, extending the intuition from previous works, that a finite width redistributes the thermal population of a resonance in mass, and the Boltzmann weight $\mathrm{e}^{-m/T}$ strongly favors the sub-pole tail of the spectral function, as illustrated for the $\rho$ in \cref{fig:spectralfunction_times_nBE}. The number of thermal parents therefore increases relative to the zero-width case, while the phase space available to their daughters shrinks. At fixed freeze-out temperature and velocity the net effect is not a simple change of normalization but a change of shape, and it differs qualitatively between species, as shown in \cref{fig:Kj_ratios}. Pions are enhanced at low $p_T$ and mildly suppressed above $\SI{1}{\GeV}$, most strongly for the S-matrix and In-medium parametrizations. Kaons are suppressed throughout, with a dip around $p_T\simeq\SI{0.5}{\GeV}$ in the In-medium case that we attribute to the melting of the $\phi$ meson, the dominant broad kaon parent. Finally, protons are enhanced across the whole range.

Since the freeze-out parameters are (re-)fitted for each spectral function scenario, part of the effect of the finite widths is absorbed into changes of the fit parameters (see \cref{tab:fitparams}). For example, including finite widths lowers the freeze-out temperature by $2$--$\SI{3}{\MeV}$ and increases the freeze-out volume $dV/dy$ by $\sim 15\%$ in all three systems.
On the other hand, the average transverse velocity $\langle\beta_T\rangle$ and the velocity profile exponent $n$ remain unchanged. However, what is not absorbed is the change in the shape of the soft pion spectrum. The fit quality improves systematically from the Dirac to the In-medium scenario, with $\chi^2/\mathrm{dof}$ dropping significantly from the on-shell to the finite-width scenarios. The residual data-to-model discrepancy at low $p_T$ is reduced accordingly, with the same ordering of the scenarios in all three collision systems, as \cref{fig:spectra_fits,fig:DatatoModel_comp} show. Nevertheless, 
even with the in-medium spectral functions the data-to-model ratio at low transverse momentum $p_T\simeq\SI{0.1}{\GeV}$ remains well above unity.
Therefore, the thermal broadening of the light vector mesons is not by itself the resolution of the soft-pion puzzle, but it must be included in any baseline computation against which novel dynamical mechanisms are compared.

There are several directions for further improvement of the current analysis.
One can include a larger set of in-medium spectral functions, which are currently available to us only for the light vector mesons. Adding that of the $\Delta(1232)$ baryon, for instance, would affect the proton spectra.
On the side of the implementation, our spectral functions are used with branching ratios evaluated at the pole mass, whereas a consistent treatment through mass-dependent partial widths would further modify the daughter spectra.
On the side of the fits, the blast-wave surface used here also neglects viscous corrections and hadronic rescattering, both of which act in the soft pion region. Embedding the finite-width feed-down in a full viscous-hydrodynamic simulation~\cite{Devetak:2019lsk,Nijs:2020ors,Lu:2024shm} is the natural next step, and the tabulated {\FastReso} decay functions make this straightforward~\cite{Kirchner:2023fsj}. Such advances would also allow for a more refined extraction of the freeze-out parameters, e.g., through a Bayesian analysis~\cite{JETSCAPE:2020mzn,Nijs:2020roc}.
We also note that the same in-medium spectral functions are used to compute the thermal dilepton spectrum, so the precision measurements of $\rho$ spectral function melting envisaged for ALICE 3~\cite{ALICE:2022wwr} can be complemented by simultaneous studies of low-momentum pion spectra.
We finally note that the increase of the pion yield also has a direct impact on the yield of decay photons~\cite{Reygers:2026pbg}, which will have an effect on the extraction of space-time information of the late stages of heavy-ion collisions through direct photons~\cite{Gotz:2021dco,Schafer:2019edr,Garcia-Montero:2019kjk}.

Resonance decays contribute around 60\% of the measured soft pions for freeze-out temperatures $T_\mathrm{fo}\sim \SI{140}{\MeV}$ (see \cref{sec:ThermalFraction}). The other 40\% are directly produced pions, which in our approach were taken to be thermal. Although small deviations from equilibrium can be accounted for by bulk and shear viscous corrections to the freeze-out spectra, there may also be large, unaccounted-for non-equilibrium corrections.
Pions are components of the order parameter of chiral symmetry breaking, which
becomes a slow degree of freedom at the chiral crossover. Therefore,
long-wavelength pions can fall out of thermal equilibrium as the QGP passes
through the QCD crossover. Recent studies of the real-time dynamics of
$O(4)$ quenches (relevant for the QCD phase transition)~\cite{Florio:2025lvu,Florio:2025zqv} indeed show an enhancement of the soft pion modes.
It is therefore the combination of the improved equilibrium baseline addressed in
this paper with the out-of-equilibrium chiral dynamics that could lead to the
resolution of the soft pion puzzle in heavy-ion collisions, and to direct
evidence of chiral symmetry restoration in QCD.

\begin{acknowledgments}

The authors would like to thank Ralf Rapp and Hendrik van Hees for sharing thermal spectral functions. 
The authors thank Jannis Gebhard, Alexander Kalweit, Pok Man Lo, Klaus Reygers, Johanna Stachel and Derek Teaney for useful discussions.
A.M. and Q.C. are supported by the DFG through Emmy
Noether Programme (project number 496831614) and CRC 1225 ISOQUANT (project number 273811115).
O.G.-M. 
was supported by the European Research Council under project ERC-2018-ADG-835105 YoctoLHC, and by Mar\'ia de Maeztu excellence unit grant CEX2023-001318-M.

The authors used Claude (Anthropic) for assistance in developing and debugging the {\FastReso} and analysis code, and for language editing of the manuscript. The authors verified all code and text and are solely responsible for the content.
\end{acknowledgments}

\appendix

\section{Numerical implementation details}\label{sec:numerical_details}

To optimize the performance of the package \FastReso{}, several changes were introduced in the new iteration. Firstly, in contrast to the original version of the code, the package only performs the decays that contribute to the selected number of final state particles, e.g., pions, kaons and protons. Secondly, the computations of \cref{eq:recursive} for different momentum bins are now parallelized with \texttt{OpenMP}.

The multidimensional integrals
are evaluated with the deterministic Cuhre cubature of the Cuba
library~\cite{Hahn:2004fe}, using the default cubature rule with relative and
absolute tolerances $\epsilon_\text{rel}=10^{-4}$ and $\epsilon_\text{abs}=10^{-12}$
and an evaluation budget capped at $10^{7}$ integrand calls per integral. The
remaining one-dimensional integrals---i.e.\ the angular integral when parent and
children are all on shell---are instead performed with the adaptive
$51$-point Gauss-Kronrod routine of GSL ($\epsilon_\text{rel}=10^{-6}$). The
multidimensional integrals were validated using the Monte Carlo Vegas integrator implemented in GSL.

\subsection{Momentum and mass grids}
\label{sec:app_grids}
In contrast to the original publication~\cite{Mazeliauskas:2018irt}, the
irreducible components are weighted with the fluid-frame momentum $|\bar \p|$,
which improves the numerical stability in the $|\bar \p|\to 0$ limit. To improve
the resolution at low momentum, the weighted components
$|\bar \p| f_{i,b}^\text{eq}$ are tabulated on a
tangent-stretched grid of fluid rest-frame momenta
\begin{equation}
  |\bar \p|_k = p_\text{ref} \tan\!\left[
     \arctan\!\left(\frac{\bar p_\text{max}}{p_\text{ref}}\right)
     \frac{k+1}{N_{\bar p}}\right]
  \label{eq:app_pbargrid}
\end{equation}
with $k=0,\dots,N_{\bar p}-1$, $p_\text{ref}=\SI{0.5}{\GeV}$,
$\bar p_\text{max}=\SI{4}{\GeV}$ and $N_{\bar p}=201$.

For particles carrying a spectral function the components depend additionally on
the off-shell mass $m_b$, and are tabulated on a uniform grid of $N_m=101$ points
covering the spectral integration window $[t_b,\, M_b + n_\text{up}\Gamma_b]$
(see \cref{sec:app_thresholds}).
Intermediate values are obtained by cubic spline interpolation in
$|\bar \p|$ for on-shell species and by a two-dimensional bicubic spline in
$(|\bar \p|, m_b)$ for broad ones. 
We use linear mass discretization because the distribution functions are
significantly populated even below the pole mass thanks to the thermal enhancement, see \cref{fig:spectralfunction_times_nBE}.

\subsection{Change of variables in the angular integral}
\label{sec:app_change_of_variables}

For numerical implementations, the angle integral in \cref{eq:recursive} is inefficient
and we apply the change of variables to improve the integral convergence.
First, the parent energy
$E(w) = m_a/m_b^2\left( E^a_{b|c}\bar E_\p - w\, p^a_{b|c}|\bar \p|\right)$ decreases
monotonically with $w$ and extends over a range
\begin{equation}
  \Delta E = E(-1)-E(1) = \frac{2\, m_a p^a_{b|c} |\bar \p|}{m_b^2},
\end{equation}
which for a light child $b$ and moderate momenta is much larger than the temperature,
$\Delta E \gg T$. Since $f^\text{eq}_{i,a}(E)\sim e^{-E/T}$, only a slice of relative
width $\sim T/\Delta E$ at the upper end $w \to 1$ of the interval contributes, making a
uniform quadrature in $w$ inefficient.  Second, explicit divisions by $m_b^2$ and $|\bar \p|$ diverge as these quantities approach zero,
although the full result remains finite.
A single integrand valid for all child masses and down to $|\bar \p|=0$ therefore requires that these cancellations be
performed analytically rather than left to floating-point arithmetic.

Both problems are solved by introducing
\begin{equation}
  y = \left[1 + (1-w)\,\frac{m_a p^a_{b|c}|\bar \p|}{m_b^2\, T}\right]^{-1},
  \label{eq:app_ydef}
\end{equation}
which maps $w\in[-1,1]$ monotonically onto $y\in[y_\text{min},1]$ with
$y_\text{min} = [1+2 m_a p^a_{b|c}|\bar \p|/(m_b^2 T)]^{-1}$. After this change of variables and some manipulations,
the kinematic functions can be written as
\begin{subequations}
\label{eq:app_EQ_y}
\begin{align}
  E &= \frac{m_a\big(m_b^{2}+(p^a_{b|c})^{2}+\bar\p^{2}\big)}
            {E^a_{b|c}\,\bar E_\p + p^a_{b|c}\,|\bar \p|}
       + T\!\left(\frac{1}{y}-1\right),\\
  Q &= \frac{m_a\big(\bar\p^{2}-(p^a_{b|c})^{2}\big)}
            {E^a_{b|c}\,|\bar \p| + p^a_{b|c}\,\bar E_\p}
       + T\!\left(\frac{1}{y}-1\right)\frac{\bar E_\p}{|\bar \p|},
\end{align}
\end{subequations}
where all $1/m_b^2$ factors have
disappeared and the only singular $m_b$ dependence sits in
$y_\text{min}$, which tends to zero as $m_b\to 0$. 
Here, it also becomes clear that $y$ measures the parent energy in units of the temperature above its kinematic
minimum, $E = E_\text{min} + T(1/y-1)$, so that
$e^{-E/T} = e^{-E_\text{min}/T}\, e^{1-1/y}$. The Boltzmann suppression is thus
resolved on an $\mathcal{O}(1)$ interval in $y$, and the same number of nodes is
adequate for all masses and momenta.

For the cubature, we introduce $u \in [0, 1]$ which rescales $y$ linearly to the unit interval via $y = y_{\mathrm{min}} + u (1 - y_\mathrm{min})$. By defining
\begin{equation}
  d_1 \equiv m_b^{2} T + 2\, m_a p^a_{b|c} |\bar \p|,
  \qquad
  d(u) \equiv m_b^{2} T + 2u\, m_a p^a_{b|c} |\bar \p|,
  \label{eq:app_Ddu}
\end{equation}
we can rewrite $y = d(u)/d_1$ and $y_\text{min}=d(0)/d_1$. The identity
$T(1/y-1) = 2T(1-u)\,m_a p^a_{b|c}|\bar \p|/d(u)$ then removes the last dangerous
ratio in \cref{eq:app_EQ_y}, because the explicit factor $|\bar \p|$ in the numerator
cancels the $\bar E_\p/|\bar \p|$ of $Q$:
\begin{subequations}
\label{eq:app_EQ_u}
\begin{align}
  E(u) &= \frac{m_a\big(m_b^{2}+(p^a_{b|c})^{2}+\bar\p^{2}\big)}
              {E^a_{b|c}\,\bar E_\p + p^a_{b|c}\,|\bar \p|}
        + \frac{2 m_a T (1-u)\, p^a_{b|c}\,|\bar \p|}{d(u)}, \\
  Q(u) &= \frac{m_a\big(\bar\p^{2}-(p^a_{b|c})^{2}\big)}
              {E^a_{b|c}\,|\bar \p| + p^a_{b|c}\,\bar E_\p}
        + \frac{2 m_a T (1-u)\, p^a_{b|c}\,\bar E_\p}{d(u)}.
\end{align}
\end{subequations}
Collecting the measure, the Jacobian
$dw/du = 2\, m_b^{2} T\, d_1/d(u)^{2}$ and the $m_a^2/m_b^2$ prefactor of
\cref{eq:recursive}, the angular
integral becomes
\begin{align}
  \frac{m_a^{2}}{m_b^{2}}\,\frac{1}{2}\int_{-1}^{1}\!dw\;
     |\bar \p^a| f_{i,a}^\text{eq}(E)\,A^\text{eq}_i
  &= \int_{0}^{1}\!\!du\; \frac{m_a^{2}\,T\,d_1}{d(u)^{2}}\nonumber\\
  &\quad\times f_{i,a}^\text{eq}\big(E(u)\big)\,
     |\bar \p^a| A^\text{eq}_i(u),
  \label{eq:app_uintegral}
\end{align}
where the explicit parent momentum $|\bar\p^a| = p(w) = \sqrt{E^2-m_a^2}$ cancels
the $1/p(w)$ of the weights $A^\text{eq}_i$.

\subsection{Mass integrations and importance sampling}
\label{sec:app_mass_integrals}

After the substitution of \cref{sec:app_change_of_variables} the remaining spectral
mass integrals in \cref{eq:recursive} are also mapped onto the interval $[0,1]$, so that the full
decay integral is a cubature over the unit hypercube of dimension $1$ (both parent and
children on shell), $2$, $3$ (broad parent and one broad child in a 2-body decay), or
$4$ (3-body decay with a broad parent and one broad child). The parent mass is mapped
linearly,
\begin{equation}
  m_a = \max\!\big(t_a,\, t_b + t_c\big)
        + \big[m_a^\text{up} - \max(t_a, t_b+t_c)\big]\,x_a,
\end{equation}
with the spectral weight $\rho_a(m_a)$ and the Jacobian retained explicitly. For the
unobserved child $c$ a uniform mapping is inefficient, because $\rho_c$ is strongly
peaked at the pole while the integration range is truncated from above by
$m_c^\text{max} = \min(m_c^\text{up},\, m_a-m_b)$, which for light parents cuts into
the peak. We therefore sample $m_c$ by inverse-transform (importance) sampling of its
own spectral function,
\begin{align}
  m_c = G_c^{-1}\!\big(G_c(m_c^\text{max})\, x_c\big),
  \quad
  G_c(m) = \frac{1}{\mathcal N_c}\!\int_{t_c}^{m}\!\! \rho_c(m')\,dm',
\end{align}
where the spectral weight collapses to the constant
$G_c(m_c^\text{max})\,\mathcal N_c$, i.e.\ the fraction of the spectral function inside
the kinematically allowed window. The cumulative distribution $G_c$ and its inverse are
tabulated once per particle on $500$ points. 

\subsection{Determination of threshold masses}
\label{sec:app_thresholds}

The lower endpoint of every mass integral in \cref{eq:recursive} is the threshold
mass $t_i$, which must be assigned consistently to every species in the decay
list. For a particle treated as on shell (Dirac, i.e.\ $\Gamma_i/M_i$ below the
cutoff $(\Gamma/M)_\text{min}=0.05$) the spectral function is a delta function and
$t_i = M_i$, its pole mass. A resonance carrying a spectral function, on the other
hand, cannot exist below the lowest invariant mass its decay products can carry,
so that
\begin{equation}
  t_i = \min\!\left(M_i,\;
        \min_{\text{channels } i\to 1\dots n}\; \sum_{k=1}^{n} t_k \right),
  \label{eq:app_threshold}
\end{equation}
where the inner minimum runs over all decay channels of $i$ listed in the decay
table and the sum over the daughters of that channel. Since the
recursion uses the daughters' thresholds and not their pole masses, thresholds
accumulate along decay chains: $t_\pi = M_\pi$ gives $t_\rho = 2 M_\pi$ from
$\rho\to\pi\pi$ and then $t_{a_1} = 3 M_\pi$ from $a_1\to\rho\pi$. 
Once $t_i$ is fixed, the support of the spectral function and the mass grid of
\cref{sec:app_grids} are initialized on $m \in [\,t_i,\; M_i +
n_\text{up}\Gamma_i\,]$ with $n_\text{up}=5$, while the spectral function is
normalized to unity over $[t_i, M_i + n_\text{norm}\Gamma_i]$ with
$n_\text{norm}=10$.

\section{Thermal fraction of the yield}
\label{sec:ThermalFraction}

It is instructive to quantify how much of the final yield is primordial. Using the
freeze-out kernels of \cref{eq:CF_kernels} computed with decay feed-down and without
(thermal), we define the differential and the integrated thermal
fractions
\begin{align}
\label{eq:thermal_fraction}
F_\text{th}(p_T,\beta_T, T_\text{fo}) &= \frac{K_1^\text{eq,th}(p_T,\beta_T,T_\text{fo})}{K_1^\text{eq,tot}(p_T,\beta_T,T_\text{fo})},
\\
\langle F_\text{th}\rangle(\beta_T,T_\text{fo}) &= \frac{\int_0^\infty dp_T\, p_T K_1^\text{eq,th}}{\int_0^\infty dp_T\, p_T K_1^\text{eq,tot}},
\end{align}
where the integrals are the $p_T$-integrated yields of \cref{eq:blast_wave_spec} for a
single fluid cell. The volume and degeneracy factors cancel in the ratio. 

\begin{figure}
    \centering
    \includegraphics[width=\linewidth]{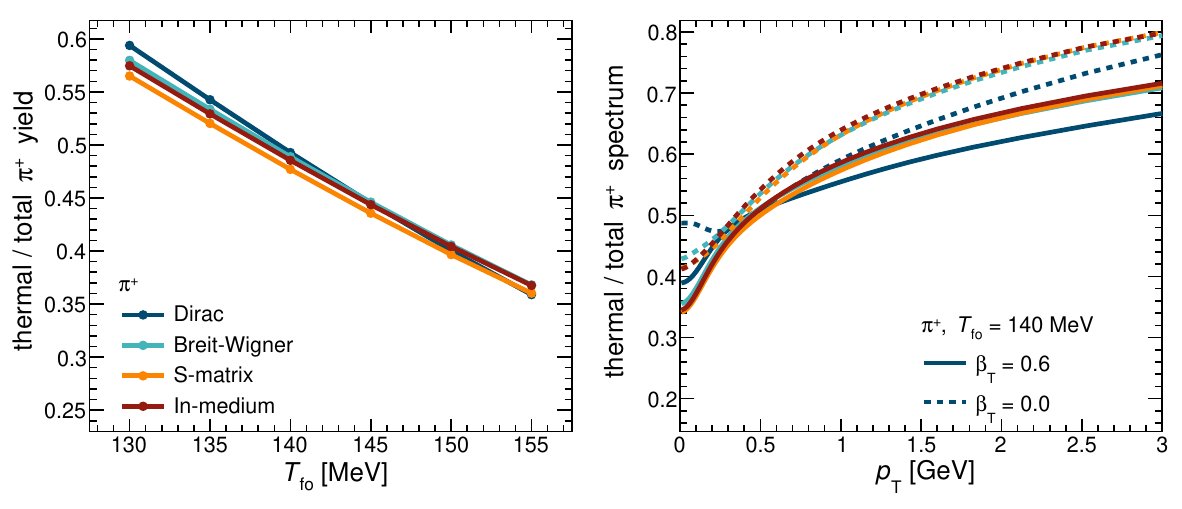}\\
    \includegraphics[width=\linewidth]{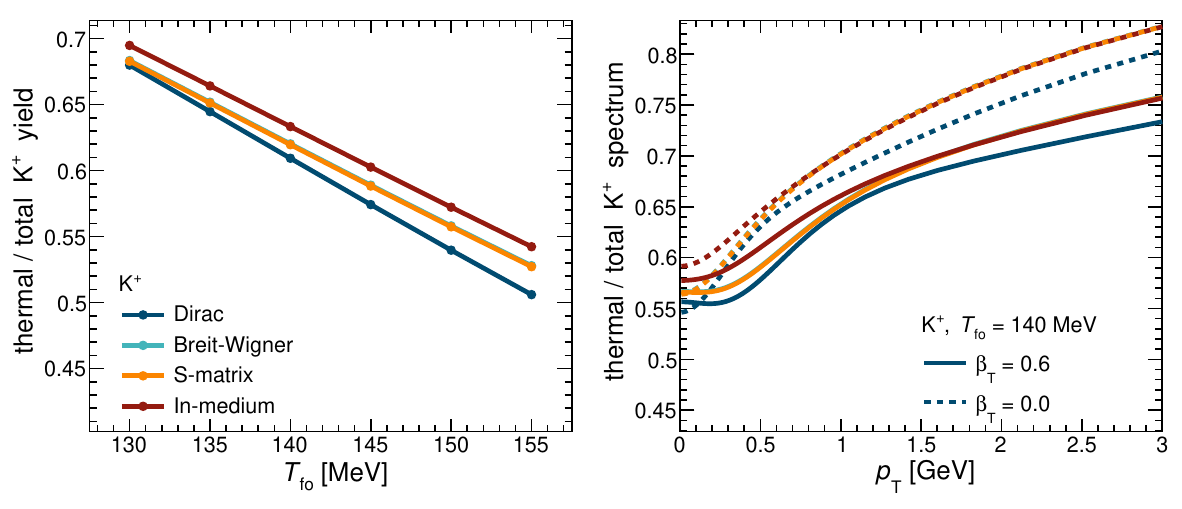}\\
    \includegraphics[width=\linewidth]{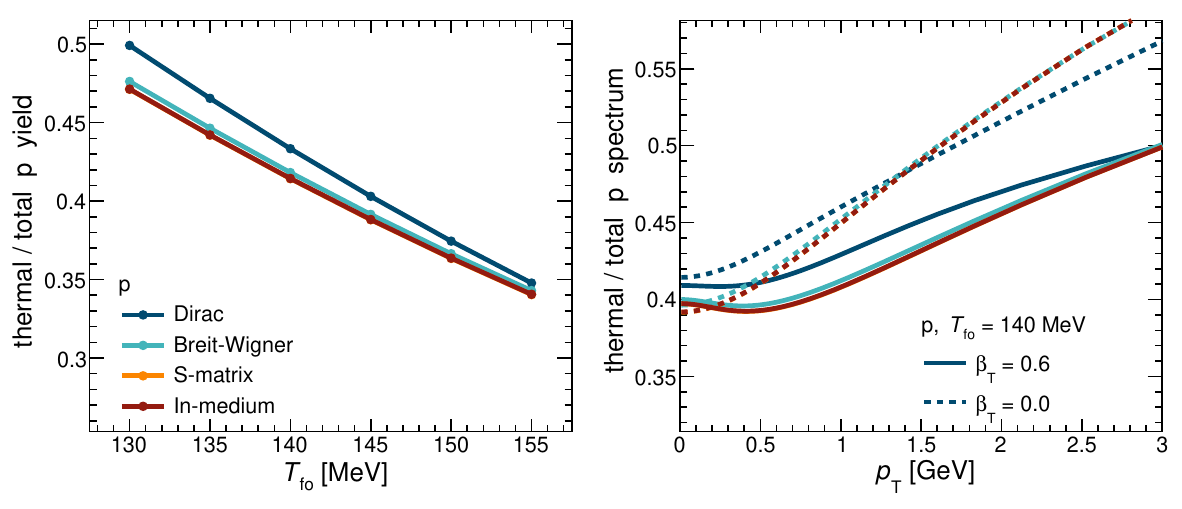}
    \caption{Thermal fraction of the yield, \cref{eq:thermal_fraction}, for $\pi^+$ (top),
    $K^+$ (middle) and $p$ (bottom) and the four spectral function scenarios. Left:
    integrated fraction $\langle F_\text{th}\rangle(\beta_T,T_\text{fo})$ as a function of the freeze-out
    temperature. Right: differential fraction $F_\text{th}(p_T)$ at $T_\text{fo}=\SI{140}{\MeV}$
    for $\beta_T=0.6$ (solid) and $\beta_T=0$ (dashed).}
    \label{fig:thermal_fraction}
\end{figure}

For a
$\tau=\text{const}$ surface the Cooper-Frye integral of a cell gives
$n\, (u\cdot d\sigma) = n\, u^\tau \tau\, dV$, so radial flow multiplies the thermal and
the total yield by the same factor $u^\tau=1/\sqrt{1-\beta_T^2}$ and only redistributes
particles in $p_T$. The integrated fraction $\langle F_\text{th}\rangle(\beta_T,T_\text{fo})$ is therefore
independent of $\beta_T$ and we
show it for $\beta_T=0$ in \cref{fig:thermal_fraction}.

The integrated fractions in the left panels of \cref{fig:thermal_fraction} fall
approximately linearly with $T_\text{fo}$, from $0.59$ to $0.36$ for pions, $0.68$ to
$0.51$ for kaons and $0.50$ to $0.35$ for protons between $T_\text{fo}=130$ and
$\SI{155}{\MeV}$: the higher the freeze-out temperature, the more heavy resonances are
excited and the larger their feed-down contribution. The sensitivity to the spectral
functions is modest but its sign is species-dependent.
For kaons the finite widths increase the thermal fraction and the ordering
(Dirac lowest, In-medium highest) is the same at all temperatures, whereas for protons
they decrease it, the Dirac scenario giving the largest fraction throughout. For pions
 the orderings interchange around
$T_\text{fo}\approx\SI{143}{\MeV}$.

The differential fractions in the right panels grow with $p_T$ for pions and kaons as
feed-down pions are concentrated at low $p_T$, where they make up about two thirds of
the yield, while for protons $F_\text{th}(p_T)$ is nearly flat below $\SI{0.5}{\GeV}$ before
rising. Radial flow lowers the differential
fraction at essentially all $p_T$, although the integrated ratio is independent of flow --- an example of Simpson's paradox. For larger $p_T$ the reduction of thermal contribution can be explained by the decay
products inheriting the larger blue shift of their heavier parents.

\bibliography{main.bib}

\end{document}